\documentclass[aps,prd,nofootinbib]{revtex4-1}
\usepackage{graphicx, amsmath, bbm, color, hyperref, pgf, tikz}

\usetikzlibrary{arrows,patterns,shapes,fit}
\definecolor{jred}{rgb}{0.8,0,0}
\definecolor{jgreen}{rgb}{0,0.7,0}
\definecolor{jblue}{rgb}{0,0,0.8}
\tikzstyle{c} =   [coordinate]
\tikzstyle{a} =	[->,line width=1pt]
\tikzstyle{vb} =	[circle, draw=black, line width=.2pt, fill=jred, inner sep=0pt, minimum size=1.5mm]
\tikzstyle{vh} =	[circle, draw=black, line width=.2pt, fill=jblue, inner sep=0pt, minimum size=1.5mm]
\tikzstyle{eb} =	[draw=jgreen,line width=.5pt]
\tikzstyle{f} = 	[fill=jblue, fill opacity=.05]

\renewcommand\[{\begin{equation}}
\renewcommand\]{\end{equation}}
\newcommand{\ba}{\begin{eqnarray}}
\newcommand{\ea}{\end{eqnarray}}

\renewcommand{\eqref}[1]{Eq.\,(\ref{#1})}

\newcommand{\vm}{\vec m}
\newcommand{\vn}{\vec n}

\newcommand{\Tr}{\mathrm{Tr}}

\newcommand{\lap}{\Delta}
\newcommand{\stm}{\mathcal{M}}		
\newcommand{\std}{D}				
\def\Ds{D_{\textsc s}}
\def\Duv{D_{\textsc s}^\alpha} 
\newcommand{\bi}{\gamma_{\textsc{bi}}}

\newcommand{\spec}{\omega^2}        
\newcommand{\size}{N_{\text{lattice}}}
\newcommand{\dof}{n}				
\newcommand{\period}{\mathcal{N}}	
\newcommand{\vx}{V}                 
\newcommand{\Acal}{\mathcal{A}}
\newcommand{\av}{\hat{\mathcal{A}}_v^{(\alpha)}}

\newcommand{\jmin}{j_{\text{min}}}
\newcommand{\jmax}{j_{\text{max}}}
\newcommand{\lmin}{l_{\text{min}}}
\newcommand{\lmax}{l_{\text{max}}}
\newcommand{\xmin}{x_{\text{min}}}
\newcommand{\xmax}{x_{\text{max}}}
\newcommand{\apt}{\alpha_*}
\newcommand{\aflow}{\alpha_*}

\def\d{\mathrm{d}}
\def\e{\textrm e}
\def\i{\mathrm i}

\def\j{\jmath}

\newcommand{\aver}[1]{\overline{#1}}  
\newcommand{\expec}[1]{\langle #1 \rangle}

\newcommand{\qbra}{\langle}
\newcommand{\qket}{\rangle}

\def\ie{{i.e.}~}

\newcommand{\COTa}{Calcagni:2013ku}
\newcommand{\COTb}{Calcagni:2014ep}
\newcommand{\COTc}{Calcagni:2015is}

\makeatletter
\def\l@subsubsection#1#2{}
\makeatother

\begin{document}

\title{Emergence of Spacetime in a restricted Spin-foam model}

\author{Sebastian Steinhaus}
\email[]{ssteinhaus@perimeterinstitute.ca}

\affiliation{Perimeter Institute f. Theoretical Physics \\ 31 Caroline St N\\
N2L 2Y5 Waterloo, ON\\
Canada}

\author{Johannes Th\"urigen}
\email[]{johannes.thuerigen@physik.hu-berlin.de}
\affiliation{Department of Physics, Department of Mathematics\\ Humboldt-Universit\"at zu Berlin\\ Unter den Linden 6\\10099 Berlin\\
Germany, EU}


\begin{abstract}
The spectral dimension has proven to be a very informative observable to understand the pro\-perties of quantum geometries in approaches to quantum gravity.
In loop quantum gravity and its spin-foam description, it has not been possible so far to calculate the spectral dimension of spacetime.
As a first step towards this goal, here we determine the spacetime spectral dimension in the simplified spin-foam model restricted to hypercuboids.
Using Monte Carlo methods we compute the spectral dimension for state sums over periodic spin-foam configurations on infinite lattices.
For given periodicity, \ie number of degrees of freedom, we find a range of scale where an intermediate spectral dimension between 0 and 4 can be found, continuously depending on the parameter of the model.
Under an assumption on the statistical behaviour of the Laplacian we can explain these results analytically.
This allows us to take the thermodynamic limit of large periodicity and find a phase transition from a regime of effectively 0-dimensional to 4-dimensional spacetime.
At the point of phase transition, dynamics of the model are scale invariant which can be seen as restoration of diffeomorphism invariance of flat space.
Considering the spectral dimension as an order parameter for renormalization we find a renormalization group flow to this point as well. 
Being the first instance of an emergence of 4-dimensional spacetime in a spin-foam model, the properties responsible for this result seem to be rather generic. We thus expect similar results for more general, less restricted spin-foam models.

\end{abstract}


\maketitle

\tableofcontents

\section{Introduction}

Any approach of quantum gravity that replaces the continuous metric for a postulated fundamental, often discrete, structure has to face the challenge to connect back to well-known continuum physics, in particular a smooth four-dimensional spacetime. Often it is argued that such a spacetime should ``emerge'' from this quantum theory in a suitable continuum limit \cite{Oriti:2014jo,Crowther:2014ut, Caravelli:2012iy, Rastgoo:2016iq}. 
While this is an appealing picture, defining and performing such a continuum limit is the essential point of ongoing research. 
Moreover, a theory of quantum gravity may allow for many different phases resulting in very different continuum spacetimes. To gain an insight and better understanding of the dynamics of such a theory, studying observables related to a notion of continuous spacetime is indispensable.

One observable ideally suited for this task is the spectral dimension $\Ds$ of spacetime, since it can be calculated both for discrete and continuous spacetimes and thus straightforwardly compared in various scenarios. In a nutshell, the spectral dimension is an effective dimension as seen by a free scalar field diffusing on the spacetime. From this diffusion process one can deduce a dimension, which in flat (continuum) spacetime agrees with the topological dimension $D$. Interestingly, this effective dimension can change depending on the length scale  at which spacetime is probed. 

In many approaches to quantum gravity such a behaviour has been observed, strikingly in a similar way \cite{Carlip:2017ik}: a flow of the spectral dimension to a value smaller than $\std=4$ occurs at short length scales \cite{ Crane:1985ba, Crane:1986gf, Ambjorn:2005fj,Ambjorn:2005fh, Coumbe:2015bq, Lauscher:2005kn,Horava:2009ho,Benedetti:2009fo,Alesci:2012jl,Arzano:2014ke, Modesto:2009bc, \COTc}.
Crane and Smolin studied the distribution of virtual black holes and found a dimensional reduction at short scales depending on the distribution \cite{Crane:1985ba, Crane:1986gf}. 
In causal dynamical triangulations (CDT) \cite{Ambjorn:2012vc} 
a dimensional flow from $\Ds=4$ to $\Ds=2$ is found at short scales in a phase resembling a de Sitter spacetime 
(though more recent CDT calculations \cite{Coumbe:2015bq} rather hint at $\Ds\simeq 3/2$).
Modified dispersion relations effect a similar dimension flow in the asymptotic safety scenario \cite{Lauscher:2005kn}, Horava-Lifshitz gravity \cite{Horava:2009ho} or non-commutative field theory \cite{Benedetti:2009fo,Alesci:2012jl,Arzano:2014ke}.
In causal set theory \cite{Dowker:aza}, a causal spectral dimension \cite{Eichhorn:2013ova} is defined essentially taking into account the causal structure fundamental to a causal set and a dimensional reduction is found at short scales \cite{Eichhorn:2017djq}. 
Also in loop quantum gravity there exist indications for a dimensional flow, either in terms of the scaling of the area spectrum \cite{Modesto:2009bc} or due to a particular superposition of spin network states \cite{Calcagni:2014uz}.
Naturally the question arises whether this behaviour is a universal feature of quantum gravity \cite{Carlip:2017ik} and what observational consequences might exist \cite{AmelinoCamelia:2013hk}.

 
\
 
Explicit evaluation of physically relevant observables like the spectral dimension remains one of the most pressing issues in the spin-foam approach to quantum gravity \cite{Perez:2013uz}.
The idea of spin-foam models is to give a rigorous definition of the quantum-gravity path integral based on the first-order formalism for general relativity as a constrained topological BF theory \cite{Plebanski:1977zz, Horowitz:1989ci}. 
There are various proposals on how two implement these  constraints on the quantum level \cite{Barrett:1998fp,Barrett:2000fr,Engle:2007em,Engle:2008ka,Engle:2008fj,Freidel:2008fv,Baratin:2012br}.
While they can be justified by giving the right semiclassical limit to the amplitudes of Regge calculus locally \cite{Barrett:2011bb}, it remains an open issue to understand and control the quantum dynamics of extended spacetimes. 
The dependence of the spin-foam amplitude on a given cell complex can be removed either by a summation rule as provided by group field theory \cite{Oriti:2014wf,Freidel:2005jy,Oriti:2006ts, Oriti:2012wt} or by a renormalization procedure based on coarse graining \cite{Dittrich:2014ui,Dittrich:2012ba,Dittrich:2012he,Dittrich:2016dc}.
Only then it is possible to define the quantum expectation value of observables.

In this work we attack the challenge to compute the spectral dimension in spin-foam models, more precisely in the well-studied Engle-Pereira-Rovelli-Livine and Freidel-Krasnov (EPRL-FK) model for 4D Euclidean gravity. Studying the spectral dimension for this model in full generality is currently out of reach due to the complexity of spin-foam models in general. Thus we restrict ourselves to a subset of the full gravitational path integral, so-called quantum cuboids \cite{Bahr:2016co, Bahr:2016dl, Bahr:2017kr}. This model encodes two major approximations. The first is a restriction of the combinatorics to hypercubic lattices. Note that this does not restrict the {\it geometry} to a hypercubic lattice as it is encoded not in the 2-complex but the group theoretic data. 
In a second step we furthermore restrict these data to be of hypercuboid form by restricting the path integral to specific coherent Livine-Speziale intertwiners \cite{Livine:2007bq} that are of cuboid form. Analogous to the calculations for the 4-simplex, the spin-foam amplitude was calculated using a stationary phase approximation, often called the large-$j$ limit, in \cite{Bahr:2016co} resulting in a rational function of irreducible representations of $\text{SU}(2)$. The associated Regge action evaluated on the stationary and critical points vanishes, which implies that the internally flat hypercuboids are glued together in a flat way resulting in flat discrete spacetime. 
Thus, quantum cuboids represent a superposition of flat discrete geometries of different shapes and sizes.

Despite its simplicity this model revealed a few interesting properties. An Abelian subgroup of diffeomorphisms which corresponds to moving an entire hyperplane of the cuboid lattice merely changes the subdivision of flat spacetime into flat building blocks \cite{Bahr:2016co}. However cuboid spin foams are not invariant under this transformation, yet for a specific choice of parameters this symmetry is almost realized. 
Furthermore the renormalization of cuboid spin foams \cite{Bahr:2016dl,Bahr:2017bn} has been studied in detail following the refinement approach via embedding maps \cite{Dittrich:2012ba,Dittrich:2013xwa,Dittrich:2014ui}. Remarkably indications for a phase transition and a UV-attractive fixed point were found for the single parameter of the model in a similar regime as for the displacement symmetry. As we see later on this regime is relevant for the spectral dimension as well.

In this article we study the spectral dimension of such quantum cuboids, first numerically and then extending the results analytically. In order to perform the numerical simulations efficiently while avoiding artifacts coming from compactness of the studied lattice, we consider $\period$-periodic spin foams, i.e.~spin foams whose labels repeat themselves after $\period$ steps in any of the four (combinatorial) directions. 
We calculate the spectrum of the Laplace operator via Fourier transform 
on an infinite spin foam, prescribed by finitely many variables.
Still, the numerical investigation is costly and cannot be extended to arbitrary large periodicities $\period$. For any finite $\period$ we find a regime (in between the minimal and maximal cut-offs) in which $\Ds$ changes continuously and can take any value between 0 and 4 (for specific values of the parameter). 

Using a conjecture on the Laplace operator inspired by \cite{Sahlmann:2010bb} we derive a general law for the spectral dimension depending on the periodicity $\period$, which is in good agreement with our numerical results. Given this relation we perform the $\period \rightarrow \infty$ limit and find a discontinuous phase transition between two phases, one given by spectral dimension $\Ds = 0$ and $\Ds = 4$, where the point of the phase transition is given by the scale-invariant spin-foam amplitude\footnote{By scale-invariant we here mean invariant under uniform scaling of all spins, i.e.~the shape of the cuboid remains the same.}. In this sense we observe the emergence of 4-dimensional spacetime. At the same time the scale-invariant amplitude hint at a potential restoration of (an Abelian subgroup of) diffeomorphism invariance, where the parameter is again in a similar regime as the UV-attractive fixed point found in the renormalization group flow.

\

This article is organized as follows: we describe the basic setup of our calculations in Sec.~\ref{Sec:QuantumCuboids} introducing the quantum cuboids, the spectral dimension, the setup of periodic spin foams and briefly discuss the numerical methods used. In Sec.~\ref{Sec:QuantumSpecDim} we present the numerical results for various periodicities and provide an analytical explanation for the results derived from a conjecture of the Laplace operator. Sec.~\ref{sec:limits} deals with the $\period \rightarrow \infty$ limit showing the discontinuous phase transition in $\Ds$ and we furthermore discuss the possibility of using the spectral dimension as a condition to define a renormalization group flow. We conclude in Sec.~\ref{Sec:Conclusion} and give a brief outlook on which models to study in the future and which qualitative features we expect to carry over.


\section{Cuboid spin foams and the spectral dimension}\label{Sec:QuantumCuboids}

In this section we provide the definition of the spectral dimension as a spin-foam observable.
We recall the general notion of spectral dimension and discuss how it can be understood as an observable in a path integral for quantum gravity.
We explain how this is defined in spin-foam gravity and derive an explicit formula in the restricted case of cuboid spin-foam dynamics. 
Finally we illustrate how we address such dynamics using $\period$-periodic configurations and explain the use of numerical methods.



\subsection{Cuboid spin-foam dynamics}

Spin-foams models provide a definition of the quantum-gravity path integral.
Here we will give a concise outline of the ideas underlying the spin-foam approach to quantum gravity and the particular restriction we use in this work.

Spin foams are defined on discrete spacetime, 
more precisely the 2-skeleton $\Gamma$ of a $4$-dimensional combinatorial (pseudo) manifold, which is a collection of $V$ vertices $v$, $E$ edges $e$ and $F$ faces $f$. This 2-complex is labelled by data from a Lie group $G$: Irreducible representations $j_f$ on the faces and intertwiners $\iota_e$ on the edges, i.e. elements in the invariant subspace of the tensor product of representations meeting at this edge. 
A spin-foam model defines an amplitude to a given labelled 2-complex by assigning an amplitude $\mathcal{A}_v$ to vertices, $\mathcal{A}_e$ to edges and $\mathcal{A}_f$ to faces. The partition function is then given as a sum over all these labellings:
\begin{equation} \label{eq:SF_statesum}
    Z = \sum_{\{\iota_e,j_f\}} \prod_f \mathcal{A}_f(j_f) \prod_e \mathcal{A}_e(\iota_e,\{j_f\}_{f \supset e}) \prod_v \mathcal{A}_v (\{\iota_e\}_{e \supset v}, \{j_f\}_{f \supset v} ) \,.
\end{equation}
For the 4D Euclidean theory, which we use in this article, the underlying symmetry group is $G=\text{Spin}(4)$, while $G=\text{SL}(2,\mathbbm{C})$ in the Lorentzian theory. The form of the amplitudes depends on the used spin-foam model.

In this article we are working with the Euclidean EPRL-FK spin-foam model \cite{Engle:2007em,Engle:2008ka,Engle:2008fj}, more precisely its generalization to arbitrary 2-complexes \cite{Kaminski:2010ba}. The idea underlying its construction is identifying $\text{SU}(2)$ representations and intertwiners with $\text{Spin}(4) \simeq \text{SU}(2) \times \text{SU}(2)$ representations and intertwiners. The identification of irreducible representations depends on the Barbero-Immirzi parameter $\bi \in \mathbbm{R} \setminus \{0, \pm 1\}$
\begin{equation}
    j_f^\pm := \frac{1}{2} | 1 \pm \bi| \; j_f \quad ,
\end{equation}
with $j_f, j_f^\pm \in \frac{1}{2} \mathbbm{N}$. 
For this map to be non-empty, the parameter $\gamma$ must be a rational number. For the rest of the article we choose $\bi < 1$.

For the intertwiners, which are essential for defining the edge and vertex amplitude of the spin-foam model, we introduce the so-called EPRL map $Y^{\bi}_e$ for each edge $e$ of the 2-complex:
\begin{equation}
    Y^{\bi}_e : \text{Inv}_{\text{SU}(2)} ( V_{j_1} \otimes \dots \otimes V_{j_n} ) \rightarrow \text{Inv}_{\text{SU}(2) \times \text{SU}(2)} ( V_{j^+_1,j^-_1} \otimes \dots \otimes V_{j^+_n,j^-_n} ) \quad .
\end{equation}
It maps an $n$-valent $\text{SU}(2)$ intertwiner, where $n$ faces carrying representations $j_f$ meet at the edge $e$, into an $\text{SU}(2) \times \text{SU}(2)$ intertwiner and consists of two parts. The map
\begin{equation}
    \beta^{\bi}_j : V_j \rightarrow V_{j^+,j^-} \quad ,
\end{equation}
is defined via the unique isometric embedding of $V_j$ into the factor of the Clebsch-Gordan decomposition of $V_{j^+,j^-} \simeq V_{j^+} \otimes V_{j^-}$. This map is applied to all faces containing that edge. To make sure the resulting vector is in the $\text{SU}(2) \times \text{SU}(2)$-gauge-invariant subspace of $V_{j^+_1,j^-_1} \otimes \dots \otimes V_{j^+_n,j^-_n}$, we project onto that subspace with the projector $\mathcal{P}$. Hence, the EPRL map is
\begin{equation}
    Y^{\bi}_e := \mathcal{P} \left(\beta^{\bi}_{j_1} \otimes \dots \otimes \beta^{\bi}_{j_n} \right) \quad .
\end{equation}
We now have all the necessary ingredients to define the amplitudes. The face amplitude is given as
\begin{equation}
    \mathcal{A}_f^{(\alpha)} := \left((2 j^+_f + 1)(2 j^-_f + 1)\right)^\alpha \quad ,
\end{equation}
which is just the dimension of the $(j^+_f,j^-_f)$ representation to the power $\alpha$. 
As proposed in \cite{Bahr:2016co}, we have introduced the additional parameter $\alpha \in \mathbbm{R}$, which turns out to be crucial in our analysis of the spectral dimension. 
To motivate it briefly, in spin foams it is commonly chosen to be either $\alpha = \frac{1}{2}$ or $\alpha = 1$. The former assigns the dimension of the $\text{SU}(2)$ spin $j_f$ to a face, the latter the respective $\text{SU}(2) \times \text{SU}(2)$ representation. 
However, beyond kinematical arguments, e.g. requiring invariance under the trivial subdivision of a face, this exponent is not fixed by dynamical arguments. 
Its choice essentially translate into a choice of the path-integral measure and is also present in other approaches to quantum gravity, e.g. Regge calculus \cite{Hamber:1999cf,Menotti:1996tm,Dittrich:2011vz}. 

The edge amplitude is simply introduced to normalize the intertwiners,
\begin{equation}
    \mathcal{A}_e := \frac{1}{\| Y^{\bi}_e \iota_e \|^2} \quad .
\end{equation}
The vertex amplitude $\mathcal{A}_v$ is given by
\begin{equation}
    \mathcal{A}_v := \text{Tr}_v \, \left( \bigotimes_{e \supset v} (Y^{\bi}_e \iota_e) \right) \quad ,
\end{equation}
where $\text{Tr}_v$ denotes the vertex trace acting on the tensor product of all intertwiners $\iota_e$ meeting at the vertex $v$ in the following way. Each face $f$ that contains the vertex $v$ is shared exactly by two edges meeting at $v$. The indices of the respective intertwiners are then contracted with one another according to the combinatorics of the 2-complex.

A spin-foam configuration has a geometric interpretation in the following way.
The intertwiners $\iota_e$ describe a quantum polyhedron, a chunk of 3D space dual to the edge $e$ whose boundary areas are given by the adjacent spins $j_f$. The combinatorics of the 2-complex then determine how these 3D chunks of space are glued together to form a 4D geometry. Summing over the areas of faces and shapes of 3D polyhedra then implements a discrete sum over 4D geometries.

Note that the 2-complex $\Gamma$ is a choice. Several possibilities have been suggested in the literature to account for the dependence on this choice. A straightforward idea is to sum over all 2-complexes of a certain class, thus capturing all possible discretizations permitting a transition between boundary states. 
The most systematic implementation of this idea is group field theory \cite{Freidel:2005jy,Oriti:2006ts,Oriti:2012wt} where spin-foam amplitudes appear as amplitudes in the perturbative sum labelled by Feynman diagrams $\Gamma$. There the question of consistency is addressed in terms of renormalizability of the theory. 
Alternatively it is possible to start with a fixed 2-complex $\Gamma$, yet one has to make sure that the calculated results are consistent with choosing a finer 2-complex $\Gamma'$. To this end one has to relate amplitudes across 2-complexes guaranteeing the same physical transitions. This is done by identifying states across boundary Hilbert spaces (via so-called embedding maps) akin to the construction of the kinematical Ashtekar-Lewandoswki vacuum in loop quantum gravity \cite{Ashtekar:1991kc,Ashtekar:1994mh}.

\subsubsection{Quantum cuboids}

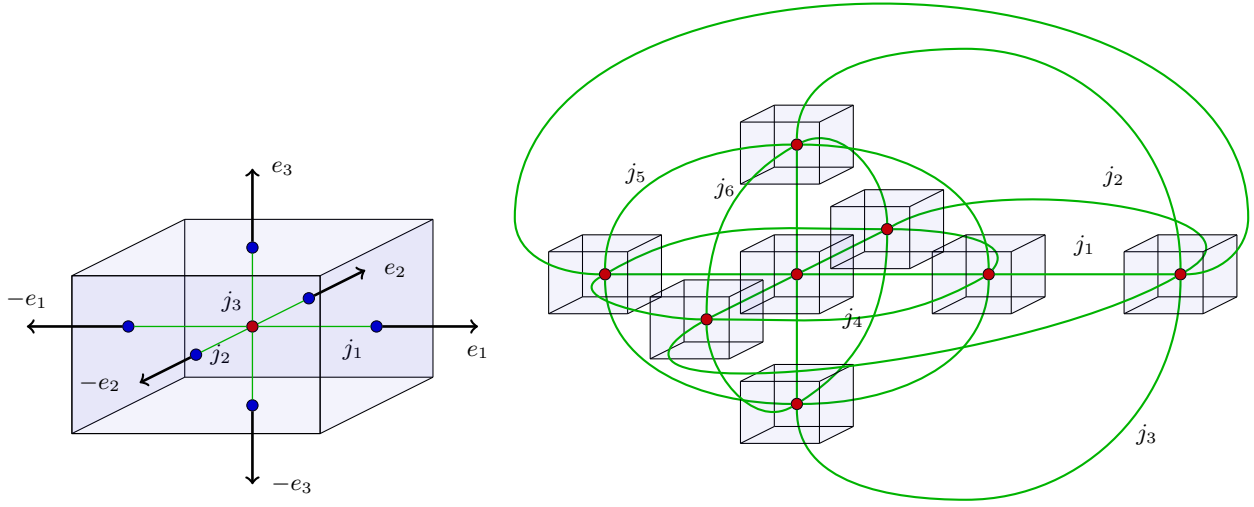
\begin{figure}
\begin{center}
\begin{tikzpicture}[scale=.75]
    \node [vb, label=100:{$j_3$}] (v) at (0,0) {};
    \node [vh] (1-) at (-2.2,0) {};
    \node [vh, label=-135:{$j_1$}] (1+) at (2.2,0) {};   
    \node [vh, label=0:{$j_2$}] (2-) at (-1,-.5) {};
    \node [vh] (2+) at (1,.5) {};
    \node [vh] (3-) at (0,-1.4) {};
    \node [vh] (3+) at (0,1.4) {};    
    \node [c] (---) at (-3.2, -1.9) {};
    \node [c] (--+) at (-3.2, .9) {};
    \node [c] (+-+) at (1.2, .9) {};
    \node [c] (+--) at (1.2, -1.9) {};
    \node [c] (+++) at (3.2, 1.9) {};
    \node [c] (-++) at (-1.2, 1.9) {};
    \node [c] (-+-) at (-1.2, -.9) {};
    \node [c] (++-) at (3.2, -.9) {};    
    \foreach \i in {+,-}{
        \foreach \j in {+,-}{
            \draw (\i\j+) -- (\i\j-);
            \draw (\i+\j) -- (\i-\j);
            \draw (+\i\j) -- (-\i\j);
        }
    }
    \foreach \i in {+,-}{
        \path [f] (\i--) -- (\i+-) -- (\i++) -- (\i-+);
        \path [f] (+\i-) -- (+\i+) -- (-\i+) -- (-\i-);
       }
    \foreach \i in {1+,1-, 2+, 2-, 3+, 3-}{
        \draw [eb] (v) -- (\i);
        }
    \draw [a] (1+) -- (4,0) node[label=below:{$e_1$}] {};
    \draw [a] (1-) -- (-4,0) node[label=above:{$-e_1$}] {};
    \draw [a] (2+) -- (2,1) node[label=right:{$e_2$}] {};
    \draw [a] (2-) -- (-2,-1) node[label=left:{$-e_2$}] {};
    \draw [a] (3+) -- (0,2.8) node[label=right:{$e_3$}] {};
    \draw [a] (3-) -- (0,-2.8) node[label=right:{$-e_3$}] {};
    \draw (---) rectangle (+-+);
\end{tikzpicture}
\tikzstyle{eb} =	[draw=jgreen,line width=.8pt]
\begin{tikzpicture}[scale=.75]
    \node [vb] (v) at (0,0) {};
    \node [vb] (+1) at (3.4,0) {};
    \node [vb] (-1) at (-3.4,0) {};
    \node [vb] (+2) at  (1.6,.8) {};
    \node [vb] (-2) at  (-1.6,-.8) {};
    \node [vb] (+3) at (0,2.3) {};
    \node [vb] (-3) at  (0,-2.3) {};
    \node [vb] (+4) at  (6.8,0) {};
    \foreach \i in {+1,-1, +2, -2, +3, -3}{
        \draw [eb] (v) -- (\i);
        }
    \draw [eb] (+1) to[out=50, in=0] (+2);
    \draw [eb] (+2) to [out=180, in=30] (-1);
    \draw [eb] (+3) to[out=25, in=90]  (+2);
    \draw [eb] (+2) to [out=-90, in=33] (-3);
    \draw [eb] (-1) to [out=-90, in=180] (-3);
    \draw [eb] (+1) to [out=-90, in=0] (-3);  
    \draw [eb] (+4) to [out=-90, in=0] node [label=right:{$j_3$}] {} (3,-4)  to [out=180, in=-90] (-3);
    \draw [eb] (+4) to [out=30, in=30] node [label=above:{$j_2$}] {} (+2);         
\foreach \i/\j in {0/0, 3.4/0, -3.4/0, 1.6/.8, -1.6/-.8, 0/2.3, 0/-2.3, 6.8/0}{
\begin{scope}[xshift=\i cm, yshift=\j cm]
    \node [c] (---) at (-1, -.7) {};
    \node [c] (--+) at (-1, .4) {};
    \node [c] (+-+) at (.4, .4) {};
    \node [c] (+--) at (.4, -.7) {};
    \node [c] (+++) at (1, .7) {};
    \node [c] (-++) at (-.4, .7) {};
    \node [c] (-+-) at (-.4, -.4) {};
    \node [c] (++-) at (1, -.4) {};
        \foreach \i in {+,-}{
        \foreach \j in {+,-}{
            \draw (\i\j+) -- (\i\j-);
            \draw (\i+\j) -- (\i-\j);
            \draw (+\i\j) -- (-\i\j);
        }
    }
    \path [f] (+--) -- (++-) -- (+++) -- (+-+);
    \path [f] (+--) -- (+-+) -- (--+) -- (---);
    \path [f] (--+) -- (-++) -- (+++) -- (+-+);
\end{scope}
}
     \draw [eb] (-1) to [out=210, in=180] (-2);
     \draw [eb] (+1) to [out=-150, in=0] node {$j_4$} (-2);    
     \draw [eb] (-3) to [out=210, in=-90] (-2);
     \draw [eb] (+3) to [out=210, in=90] node [label=above:{$j_6$}] {} (-2);
     \draw [eb] (-1) to [out=90, in=180] node [label=left:{$j_5$}] {} (+3);
     \draw [eb] (+1) to [out=90, in=0] (+3);         
     \draw [eb] (+4) --  node [label=above:{$j_1$}] {} (+1); 
    \draw [eb] (+4) to [out=0, in=-90] (8,1) to [out=90, in=90] (-5,1) to [out=-90, in=180] (-1);
     \draw [eb] (+4) to [out=90, in=0] (3,4) to [out=180, in=90] (+3);
    \draw [eb] (+4) to [out=210, in=210] (-2);
\end{tikzpicture}

\end{center}
\caption{Left: The six-valent cuboid intertwiner. Spins $j_i$ on opposite edges of the vertex are the same and the corresponding unit vectors $\pm e_i$ are antiparallel. 
In the semiclassical limit the spins are the areas and the unit vectors the normals of the cuboid faces.
Right: The spin network dual to a hypercuboid. Because of the translation invariance of spins along each cuboid's face direction in $\std=4$ there are $\binom{D}{2}=6$ spins.
\label{fig:cuboid}}
\end{figure}

In the present work we study the spectral dimension of spin foams restricted to hypercuboid geometries \cite{Bahr:2016co}.
We choose the combinatorial 2-complex $\Gamma$ to be hypercubic in the sense that it is the 2-skeleton of (the combinatorial dual complex of) a hypercubic lattice. 
In general, such a choice does not imply that also the geometry is hypercubic as this is encoded by the group representation data labelling the foam. 
To specify hypercuboid geometries, we consider the state sum \eqref{eq:SF_statesum} not for all possible intertwiners $\iota_e$ but only for a specific one which is sharply peaked on the shape of a cuboid. 
We define a cuboid intertwiner $\iota_{j_1, j_2, j_3}$ as the 6-valent coherent Livine-Speziale intertwiner \cite{Livine:2007bq}
\begin{eqnarray}\label{Eq:QuantumCuboid}
|\iota_{j_1,j_2,j_3}\rangle\;=\;\int_{SU(2)}dg\;g\triangleright\bigotimes_{i=1}^3|j_i,e_i\rangle|j_i,-e_i\rangle \, ,
\end{eqnarray}
that is the group-averaged tensor product of six coherent $\text{SU}(2)$ states peaked on the directions given by the (ortho\-gonal) unit vectors $e_1 = \exp(-i \pi \sigma_2 /4) \triangleright e_3$, $e_2 = \exp(-i \pi \sigma_1 /4) \triangleright e_3$ and $e_3$ in $\mathbbm{R}^3$. Note that the spins on opposite faces are chosen to be equal and the associated normal vectors are anti-parallel. The normal vectors $e_i$ assigned to adjacent faces are orthogonal matching the cuboid geometry (see Fig.~\ref{fig:cuboid}). 
This intertwiner exists for all choices of spins $j_1$, $j_2$ and $j_3$ and is always non-vanishing.

The cuboid intertwiner defines the amplitudes of this restricted EPRL model. 
First the intertwiner is boosted by the previously introduced EPRL map $Y^{\bi}_e$. For a Barbero-Immirzi parameter $\bi<1$ the vertex amplitude factorizes as $\mathcal{A}_v=\mathcal{A}_v^+\mathcal{A}_v^-$, i.e. each of them only depends on the spins $\{j^+\}$ or $\{j^-\}$ respectively. These amplitudes are given by
\begin{eqnarray}\label{Eq:AmplitudeIntegral}
\mathcal{A}_v^\pm = \int_{SU(2)^8}dg_a\;e^{S^\pm[g_a]}
\end{eqnarray}
\noindent with the complex action
\begin{eqnarray}\label{Eq:ComplexAction}
S^\pm[g_a]\;&=&\;\frac{1\pm\gamma}{2}\sum_l2j_l\ln\langle-\vn_{ab}|g_{a}^{-1}g_{b}|\vn_{ba}\rangle \quad , 
\end{eqnarray}
where $a$, $b$ denote the intertwiners, $l$ the (oriented) links connecting them in the vertex trace and $| \vec{n}_{ab} \rangle$ the $\text{SU}(2)$ coherent state peaked on the directions given by $\vec{n}_{ab}$ in the fundamental representation $j=\frac{1}{2}$. 

For the remainder of the article, we will work in the large-$j$ limit, usually understood as the semiclassical limit when keeping areas $A_f\sim \hbar j_f$ fixed, which was computed first for the EPRL-FK model for boundary data corresponding to a generic, non-degenerate 4-simplex \cite{Conrady:2008mk,Barrett:2009ci,Barrett:2009mw}.
This limit allows us to perform a stationary-phase analysis of the vertex amplitude \eqref{Eq:AmplitudeIntegral} by simultaneously scaling up all spins at once. For hypercuboid intertwiners the calculation was performed in \cite{Bahr:2016co}, and here we only review the basic steps and present the results.

The vertex amplitude $\mathcal{A}^\pm_v$ is given as an integral over eight copies of $\text{SU}(2)$, where one integration is obsolete due to the invariance of the Haar measure. The remaining integral has several isolated critical points which are related by the $2^7$-fold symmetry under changing $g_a \rightarrow -g_a$. Modulo this symmetry the two critical and stationary points satisfy the following equations for all links $(ab)$,
\begin{equation}
    g_a \triangleright \vec{n}_{ab} = -g_b \triangleright \vec{n}_{ba} \quad .
\end{equation}
The two distinct solutions and more details can be found in \cite{Bahr:2016co}. The solution reads
\begin{eqnarray}
\mathcal{A}^\pm_v\;=\;\left(\frac{1\pm\bi}{2}\right)^{\frac{21}{2}}\mathcal{B}_v
\end{eqnarray}

\noindent with
\begin{eqnarray}\label{Eq:TwoStationaryCriticalPoints}
\mathcal{B}_v(j_1,\ldots,j_6)\;=\;\frac{1}{\sqrt{\det H}}\;+\;\text{c.c.}
\end{eqnarray}
$H$ denotes the matrix of second derivatives of $S$, and its determinant is given by
\begin{eqnarray}
\det H &=& 2 \prod_{\underset{\{1256\},\{1346\},\{2345\}}{\{ijkl\}=}} \left(\sum_{\{abc\}\subset\{ijkl\}}j_a j_b j_c\right)
    \prod_{\underset{\{124\},\{135\},\{236\},\{456\}}{\{ikl\}=}}\left( (1+\i) j_i j_k j_l + \sum_{\{ab\}\subset\{ikl\}}(j_a^2 j_b + j_a j_b^2)\right) 
\end{eqnarray}
where 
the first product is over all but two opposite edges in a tetrahedron with edges labelled 1,2,3,4,5,6 as is a multiple subgraph in the hypercuboid spin network (Fig.~\ref{fig:cuboid}), the second product runs over the tetrahedron's faces.
Note that the action $S^\pm$ evaluated on the critical and stationary points vanishes exactly. This shows, at least in the large-$j$ limit, that the dynamics demand the quantum cuboids to be glued together in a flat way. Furthermore we can readily see that the vertex amplitude is a purely rational function and does not contain any oscillating parts.

In a similar way one calculates the asymptotic expansion of the edge and face amplitudes. Here we simply present the results,
\begin{equation}
    \| Y^{\bi}_e  \iota_{j_1,j_2,j_3} \| \sim \frac{8 (1-\bi^2)^{-\frac{3}{2}}}{(j_1 + j_2) (j_1 + j_3) (j_2 + j_3)} \quad ,
\end{equation}
and
\begin{equation}
    \mathcal{A}_f^{(\alpha)} \sim  j_f^{2 \alpha} \quad .
\end{equation}

Combining these results, we obtain the partition function \eqref{eq:SF_statesum} in the large-$j$ limit,
\begin{eqnarray}\nonumber
Z\;&\sim&\;\left(\frac{1-\gamma^2}{4}\right)^{\alpha F-\frac{3}{2}E+\frac{21}{2}V}\sum_{j_f}\prod_fj_f^{2\alpha}\prod_e(j_1+j_2)(j_2+j_3)(j_1+j_3)\prod_v \mathcal{B}_v^2 \\[5pt]\label{Eq:AsymptoticStateSum}
&=:&\;\left(\frac{1-\gamma^2}{4}\right)^{(6\alpha-9/2)V}\sum_{j_f}\prod_v\av \quad .
\end{eqnarray}
In the last step we have combined the vertex, edge and face amplitudes into single amplitude $\av$ assigned to each vertex thanks to the regular combinatorics of the hypercubic lattice.

%
%
Let us briefly recall some properties of the amplitude $\av$. 
It is a homogeneous function of degree $12 \alpha - 9$ in all spins, 
\[\label{eq:nongeometric-scaling}
\av(\{\lambda j_i\})\;=\;\lambda^{12\alpha-9} \av(\{j_i\}) \quad .
\]
This scaling property plays an important role later on.
Furthermore, $\av$ depends on six $\text{SU}(2)$ spins $j_i$ giving the area of the faces of the hypercuboid. These are two more degrees of freedom than the four edge lengths prescribing a hypercuboid allowing for non-geometric configurations. Non-geometricity here refers to a non-matching of shapes of faces \cite{Dona:2017dvf}, reminiscent of the twisted geometries parametrization of the loop quantum gravity phase space \cite{Freidel:2010eb}. 


\

For hypercuboid spin foams to be purely geometric, the six spins prescribing it have to satisfy three conditions:
\begin{equation} \label{eq:volume_constraint}
j_{1} \, j_{6} = j_{2} \, j_{5} = j_{3} \, j_{4} \quad ,
\end{equation}
which state that the $4$-volume calculated from these spins is the same no matter which two faces (only sharing a vertex in the dual discretization) of the hypercuboid are chosen. Then the areas of faces can be unambiguously expressed in terms of four edge lengths. Note that if two of the above equations are satisfied the third follows automatically\footnote{Each 3D cuboid is prescribed by three areas, which are in one-to-one correspondence to three edge lengths unless one of the areas vanishes. \eqref{eq:volume_constraint} ensures that these edge lengths from individual cuboids agree for all cuboids of the hypercuboid, ensuring shape matching of faces.}. These conditions are closely related to volume simplicity constraints \cite{Belov:2017who}.

Restricting the spin-foam state sum to geometric configurations implies integrating over edge lengths instead of $\text{SU}(2)$ spins. Besides the Jacobian picked up due to the change of variables, we have to take into account that we integrate over a submanifold of the original integration domain. The necessary Fadeev-Popov determinant taking care of the additional spins being gauge fixed is derived in detail in Appendix B of \cite{Bahr:2017kr}.

After this restriction the only configurations that are allowed are irregular lattices, yet all angles are right angles, i.e. the hypercuboids can take any shape as long as they remain hypercuboids. As a result all internal deficit angles vanish, such that these geometries are flat. Thus non-trivial effects on the spectral dimension of the quantum geometry can only stem from the way these geometries are superposed. In turn this is solely determined by the weight assigned to them by the spin-foam amplitudes, in particular geometries of different {\it scale}.

The function $\av$ as a function of edge lengths has a different scaling behaviour compared to the spin case due to the restriction to purely geometric geometries (encoded in the Fadeev-Popov determinant). It is still a homogeneous function in all edge lengths, but of degree $24 \alpha -14$:
\[\label{eq:geometric-scaling}
\av(\{\lambda l_i\})\;=\;\lambda^{24\alpha-14} \av(\{l_i\}) \quad .
\]
Again, this behaviour becomes important in our analysis of the results below.


\subsection{Spectral dimension and Laplacian on spin-foam geometry}\label{Sec:QuantumSpecDim}

The spectral dimension on a geometry is the dimension as seen by a fictitious field propagating on that geometry.
The standard case is a Riemannian manifold $(\stm,g)$ where the spectral dimension is defined as the scaling
\[
\Ds(\tau):=-2\frac{\partial\ln P(\tau)}{\partial\ln\tau}\;.
\]
of the trace
\[\label{eq:heat-trace}
P(\tau) = \Tr_\stm K(x,x_0;\tau) = \int_\stm \d x_0\, \sqrt{g}\, K(x_0,x_0;\tau)
\]
of the heat kernel $K(x,x_0;\tau)$ solving
\[\label{eq:heat-equation}
\partial_{\tau}K(x,x_0;\tau)-\Delta_{x}K(x,x_0;\tau)=0
\]
with appropriate boundary conditions (usually $K(x,y;0)=\delta(x-y)$).
Thus, the heat-kernel trace $P(\tau)$ depends on the geometry via the Laplacian operator $\Delta$ acting on functions, \ie scalar fields $\phi$.

We consider the spectral dimension of quantum spacetime as the scaling of the quantum expectation value of the heat-kernel trace and focus on path-integral formalism here.
Conceptually, suppose there is a definition of a gravity path integral as a sum over geometries $g$
\[
Z = \int_\mathcal{M} \mathcal{D}g \, \e^{\i S_\textsc{GR}[g]} \, .
\]
Then, the according quantum spectral dimension should be defined in terms of the expectation value of the heat kernel as an insertion into the path integral,
\[\label{eq:heat-trace-expectation}
\qbra P(\tau) \qket = \frac1{Z} \int_\mathcal{M} \mathcal{D}g \, P(\tau)\, \e^{\i S_\textsc{GR}[g]}\, ,
\]
that is 
\[\label{eq:spectral-dimension}
\Ds = -2\frac{\partial\ln \qbra P(\tau) \qket}{\partial\ln\tau}\;.
\]

Here we consider the precise definition of a path-integral proposal as given by the spin-foam dynamics defined in the previous section.
This is a discrete path integral.
Thus the main remaining task for the definition of the quantum spectral dimension is the definition of the Laplacian in the heat equation \eqref{eq:heat-equation} on spin-foam configurations.
Then the heat kernel and its trace $P(\tau)$, the observable to be inserted in the state sum, follow as solutions to this equation.




\

The Laplace operator on spin configuration can be defined using a proper definition on discrete geometries.
The Laplacian acting on a function $\phi$ defined on the dual vertices $v_n$ of a four-dimensional combinatorial complex with attached geometry depends on the four-volumes $V_{(4)}^{n}$ dual to $v_n$, 
the boundary three-volumes $V^{m n}$ between dual vertices $v_n$ and $v_m$ as well as 
the lengths $l_\star^{m n}$ of the edges dual to these three-volumes \cite{\COTa}:
\begin{eqnarray}\label{discrete-laplacian}
-(\Delta \phi)_m &=& -\sum_{n\sim m} \Delta_{mn} (\phi_m - \phi_n)\nonumber\\
&=& \frac1{V_{(4)}^m} \sum_{n\sim m} \frac{V^{mn}} {l_\star^{mn}} (\phi_m - \phi_n)\,,
\end{eqnarray}
where $n\sim m$ indicates adjacency of the vertices $v_m,v_n$.
There are various ways to define these volumes and dual lengths in terms of spin-foam degrees of freedom \cite{\COTa}.
Area variables are most natural as they are directly related to the spins $j$ labelling the configurations \cite{Livine:2009bz}, but they might be insufficient to uniquely determine discrete geometry \cite{Barrett:1999fa}. 
This issue can be overcome using flux or area-angle variables \cite{Baratin:2011hc, Dittrich:2008hg}. 
For the present purpose area variables given by spins are sufficient because all angles are considered as right angles in the  hypercuboid restriction.

\

The Laplacian on cuboid spin foams can be expressed in terms of spins.
A {semiclassical configuration} has a geometric interpretation in terms of  an assignment of areas 
\[\label{eq:area-spectrum}
A_f = \ell_\gamma^2 j_f 
\]
to the squares of the hypercubic lattice (where the dimensionfull constant $\ell_\gamma$ is of the order of the Planck length and might depend further on the Barbero-Immirzi parameter $\bi$).
Due to the form of cuboid intertwiners \eqref{Eq:QuantumCuboid}, two areas agree whenever they are parallel and one is reached from the other by a translation perpendicular to them.
That is, denoting directions on the lattice $\mu,\nu,...\in\{0,1,2,3\}$ and lattice sites $\vn\in\mathbbm{Z}^4$,
\begin{eqnarray}\label{Eq:LatticeSymmetry}
A_{\mu\nu}^{\vn + r e_\rho+ s e_\sigma} = A_{\mu\nu}^{\vn}
\end{eqnarray}
where $\mu,\nu,\rho,\sigma$ are all different directions,  $e_\mu$ are unit vectors in the $\mu$ direction, and $r,s\in\mathbbm{Z}$.

Areas uniquely determine 3d cuboid geometry but not 4d hypercuboid geometry.
A cuboid is equally determined by its three edge lengths $l_1, l_2, l_3$ or square areas $A_{ij}=l_i l_j$ in terms of the inverse relation
\[\label{eq:ltoA}
l_i^2 = \frac{A_{ij} A_{ik}}{A_{jk}} \, .
\] 

A {semiclassical quantum hypercuboid} dual to a vertex $\vn\in\Gamma$ is the 4d geometry determined by the six areas $A_{\mu\nu}^{\vn}$, $\mu<\nu$.
It has
two extra degrees of freedom compared to the four edge lengths $l_\mu$  of classical hypercuboid in $\mathbbm{R}^4$.
Accordingly, it is generically not geometric in the sense that using edge-area relations \eqref{eq:ltoA} on each cuboid face does not lead to a consistent set of edges, an instance of ``twisted geometry" \cite{Freidel:2010eb}. 
This is only the case if the geometricity conditions \eqref{eq:volume_constraint} are fulfilled, that is for areas
\begin{eqnarray}\label{Eq:GeometricityConstraints}
A_{01}A_{34}=A_{13}A_{24}=A_{14}A_{23} \,.
\end{eqnarray}
The three terms are simply the possible expressions of the 4-volume in terms of areas. 
One can check that if these agree, all expressions for edge lengths agree as well.

Edge lengths and  4-volumes of a semiclassical quantum hypercuboid can be defined naturally as averages over their distinct expressions in area.
Thus, the generalized 4-volume of such a hypercuboid dual to $\vn$ is
\begin{eqnarray}\label{eq:4Volume}
V_{(4)}^{\vn} :=\left({\prod_{\mu<\nu} A^{\vn}_{\mu\nu}}\right)^{\frac1 3} \, .
\end{eqnarray}
This allows to define the generalized length of an edge in direction $\mu$ in this hypercuboid as
\begin{eqnarray}\label{eq:length}
l^{\vn}_\mu := \frac{V_{(4)}^{\vn}}{(A^{\vn}_{\nu\rho}A^{\vn}_{\nu\sigma}A^{\vn}_{\rho\sigma})^{\frac{1}{2}}} = \frac{(A^{\vn}_{\mu\nu}A^{\vn}_{\mu\rho}A^{\vn}_{\mu\sigma})^{\frac{1}{3}}}{(A^{\vn}_{\nu\rho}A^{\vn}_{\nu\sigma}A^{\vn}_{\rho\sigma})^{\frac{1}{6}}} \, .
\end{eqnarray}
With these definitions one has $V_{(4)} = l_0 l_1 l_2 l_3$ even for non-geometric configurations.

Similar to the generalized 4-volume, \eqref{eq:4Volume}, we define 3D volumes for neighbouring vertices $\vm$ and $\vn = \vm + e_\mu$
\[
V^{\vm\vn} := V^{\vm}_{\nu\rho\sigma} 
=  l^{\vm}_\nu l^{\vm}_\rho l^{\vm}_\sigma 
=   \left(A^{\vm}_{\nu\rho} A^{\vm}_{\nu\sigma}A^{\vm}_{\rho\sigma} \right)^{\frac1 2} 
\]
using that areas on the boundary between $\vm$ and $\vn$ match due to the translation invariance \eqref{Eq:LatticeSymmetry}.
Finally, a simple way to define the dual lengths $l_\star$ is in terms of a geometric mean: 
\[
l_\star^{\vm \vn} = \sqrt{l^{\vm}_\mu l^{\vn}_\mu} 
\overset{\textrm{\eqref{eq:length}}}{=} 
\frac{(A^{\vm}_{\mu\nu}A^{\vm}_{\mu\rho}A^{\vm}_{\mu\sigma})^{\frac{1}{6}}}{(A^{\vm}_{\nu\rho}A^{\vm}_{\nu\sigma}A^{\vm}_{\rho\sigma})^{\frac{1}{12}}}
\frac{(A^{\vn}_{\mu\nu}A^{\vn}_{\mu\rho}A^{\vn}_{\mu\sigma})^{\frac{1}{6}}}{(A^{\vn}_{\nu\rho}A^{\vn}_{\nu\sigma}A^{\vn}_{\rho\sigma})^{\frac{1}{12}}} \, .
\] 
With these definitions, the matrix elements of the Laplacian as functions of areas are
\[\label{eq:area-Laplacian}
\Delta_{\vm\vn} = \frac1{\sqrt{A^{\vm}_{\mu\nu}A^{\vm}_{\mu\rho}A^{\vm}_{\mu\sigma}}}
 \frac{(A^{\vn}_{\nu\rho}A^{\vn}_{\nu\sigma}A^{\vn}_{\rho\sigma})^{\frac{1}{3}}}{(A^{\vn}_{\mu\nu}A^{\vn}_{\mu\rho}A^{\vn}_{\mu\sigma})^{\frac{1}{6}}}\, .
\]
With this Laplacian, the heat equation \eqref{eq:heat-equation} becomes a difference equation solved by the heat kernel $K(\vn,\vm;\tau)$ on the semiclassical spin-foam configuration. 
Its trace analogous to \eqref{eq:heat-trace} is simply the sum over vertices in the 2-complex $\Gamma$ \cite{\COTa,Thurigen:2015uc},
\[
P(\tau) = \Tr K(\vn,\vm;\tau) = \sum_{\vn\in\Gamma} K(\vn,\vn;\tau) \, .
\]
This completes the definition of the spectral dimension on a single semiclassical spin-foam configuration as the scaling \eqref{eq:spectral-dimension} of $P(\tau)$.
Inserting this expression in the spin-foam state sum yields corresponding quantum expectation value.

While the given definitions of semiclassical geometry entering the Laplacian, though natural, might allow also for alternatives, results on the spectral dimension of discrete geometries \cite{\COTb, Thurigen:2015uc} suggest that it is not sensitive to details of the precise definition of local geometry.
For example, from the perspective of discrete geometry of the dual complex, a dual-length definition in terms of the arithmetic mean $l_\star^{\vm \vn} = (l^{\vm}_\mu + l^{\vn}_\mu)/2$ might be more appropriate \cite{\COTa}.
While the resulting expression for Laplacian coefficients $\Delta_{\vm\vn}$ is slightly more involved, this is irrelevant for all practical purposes.
In particular, we have cross-checked that the calculations presented in this paper with either definition show no significant differences.


\subsection{Approaching the full dynamics in terms of periodic configurations}\label{sec:periodic-configurations}

Evaluating the quantum spectral dimension remains an intriguing challenge even after restricting the spin-foam path integral to the asymptotic regime of quantum cuboids \eqref{Eq:AsymptoticStateSum}. This challenge is posed by both ingredients of the calculation, the spectral dimension and computing the spin-foam state sum. To clarify this point, let us disentangle the two and discuss first the evaluation of the spectral dimension of one lattice.

Consider a discrete geometry, for simplicity with periodic boundary conditions, i.e. a $D$-dimensional torus, on which the discrete Laplace operator \eqref{discrete-laplacian} is well defined. Its spectral dimension can be reliably determined in a certain regime of the diffusion time $\tau$, namely $a \ll\sqrt{\tau} \ll a \size$. Here $a$ denotes the lattice scale and $\size$ is the number of lattice sites in each direction.%
\footnote{For simplicity we assume the lattice to be of equal size in all dimensional directions.} If $\tau$ is smaller or similar to the lattice scale $a$, the return probability remains constant or only changes slightly as we cannot resolve the geometry below the lattice scale, resulting in a spectral dimension $\Ds=0$. Seen by a random walker the diffusion time is too small for the random walker to explore the surrounding geometry or for it to even leave the initial lattice site. Conversely when $\tau$ goes beyond the size $\size$ of the lattice the return probability becomes constant again due to the compactness of the geometry, and thus $\Ds=0$. Again for a random walker the diffusion time was long enough for the walker to travel through the entire geometry to arrive back at the starting point. Hence in order to observe a non-trivial spectral dimension the lattice size must be large enough, which is determined by the number of lattice sites in each direction. Typically $\size$ should be at least of the order $\sim 10^3$. However it might be necessary that the number of lattice sites is a few orders of magnitude larger than this such that the compactness is not seen too early. An example could be a 2-torus with a small and a large radius: If the lattice scale is too large, one cannot resolve the small radius.

Given such a lattice, for concreteness in $\std=4$, it is numerically challenging to compute the spectral dimension. In order to derive the return probability at all scales $\tau$, in particular at the intermediate ones between the lattice and compactness scale, we have to know the {\it entire} spectrum of the Laplace operator $\Delta$. For $\size \sim 10^3$ in each direction, $\Delta$ is a $10^{12} \times 10^{12}$ matrix that needs to be diagonalized. Already memory cost in defining such a matrix is very high, not to mention the computational cost in computing its complete spectrum.%
\footnote{Since $\Delta$ is proportional to the adjacency matrix many of its entries are empty. 
Thus one might be tempted to work with sparse matrices instead.
However, algorithms computing eigenvalues of sparse matrices are only efficient for small number of eigenvalues, usually its largest or smallest. Such an approach is hence only feasible when studying small or large scales respectively.}

Instead of diagonalizing the entire matrix one can study the return probability of a discrete geometry via a random walker, similar to the studies in causal dynamical triangulations \cite{Ambjorn:2012vc}. The random walker randomly jumps from lattice site to lattice site, where the jump probabilities are related to the entries of $\Delta$. The return probability literally is the probability of the random walker to return to the lattice site it started from after $\sigma$ steps, where $\sigma$ can be related to the diffusion scale $\tau$. This probability is then averaged over each lattice site as a possible starting point of the random walker. Particular care must be given to the implementation of the algorithm as soon as cells can vary in their respective volume, which we detail in a companion article. In order to correctly estimate the return probability a random walk must be frequently repeated. While it is very efficient in memory consumption and thus can be straightforwardly implemented, the number of possible paths grows exponentially requiring a similarly growing number of repetitions to allow for accurate results. Hence the computational cost grows quickly when large lattice distances are considered. Since each random walk is independent of the others, this process is straightforwardly parallelizable.

Conversely the formal requirements to accurately calculate the spectral dimension pose a serious challenge to the spin-foam approach. So far most studies of spin-foam models did not exceed a few spin-foam vertices, in fact most calculations were actually performed for a single 4-simplex in the so-called large-$j$ limit. Going beyond this asymptotic expansion has recently been explored in \cite{Dona:2017dvf}, see also \cite{Christensen:2001eu} concerning the Barrett-Crane spin-foam model \cite{Barrett:1998fp,Barrett:2000fr}, but it has not yet been used to study the spin-foam partition function for discretizations consisting of several building blocks. One of the authors studied the restricted partition function of hypercubic spin foams for slightly larger discretizations in the large-$j$ limit in \cite{Bahr:2016dl,Bahr:2017kr}. However at the current stage studying the spectral dimension of spin foams in full generality is out of reach.

To make matters worse, a dynamical quantum geometry might require even larger lattice if one intends to study the quantum spectral dimension. Without imposing restrictions onto the spin-foam state sum, we superimpose geometries of varying size since we are summing over spins, which can range over many orders of magnitude\footnote{Whether this actually is the case and which geometries are preferred depends crucially on the spin-foam amplitudes.}. 
Hence it is straightforward to conceive a scenario where we superimpose a small and a comparatively large geometry. In a foam with a fixed number of vertices this can result in a situation where we run into the compactness scale of the small geometry, which can result in a biased result.

Naturally the question arises how we can reconcile the need of studying small lattice in order to keep the spin-foam state sum tractable with the imperative to make to the lattice as large as possible (if not infinite) in order to reliably compute the spectral dimension. To kill two birds with one stone we introduce $\period$-periodic lattice, i.e. we introduce an internal $\period$-periodicity to the lattice: after $\period$ steps the geometric labels of the foam are the same, in any of the four dimensions. On the one hand this greatly simplifies the calculation of the spectrum of the Laplacian: instead of diagonalizing a matrix of the size of the entire lattice we perform a Fourier transform and calculate the spectrum of the Brillouin zone from a $\period^D \times \period^D$ matrix. The total lattice size then merely determines which discrete frequencies of this spectrum are allowed. In the limit of an infinite lattice we simply integrate over the entire spectrum when calculating the return probability.

On the other hand, the spin-foam state sum gets drastically simplified, since the dynamical variables have to respect the $\period$-periodicity. Hence the summation is only over the variables associated to an `$\period$-cell'. This $\period$-cell then determines the geometry of the entire lattice, where the configuration of this cell is weighted by the spin-foam amplitudes of the entire lattice. Increasing the total lattice site implies weighing the configuration with a higher power of spin-foam amplitudes. The latter fact makes numerical simulations increasingly difficult and thus obstructs us from taking the limit of large lattices, such that we have to introduce an additional approximation: We assume the $\period$-cell to be weighted only by the spin-foam amplitudes associated with that $\period$-cell. Let us justify this approximation:

As we have discussed above the $\period$-periodicity greatly simplifies our calculation: the {\it entire} spectrum is readily available and the spin-foam state sum is much more tractable. The choice to take the infinite lattice limit is solely introduced to avoid the compactness scale of some quantum geometries and does not qualitatively affect the spectrum otherwise.
Furthermore increasing the power of the spin-foam amplitudes changes the behaviour only slightly, which we will discuss below in more detail. The crucial limit indispensable for computing the spectral dimension is taking $\period \rightarrow \infty$. Using a conjecture about the scaling behaviour of $\Delta$ inspired by \cite{Sahlmann:2010bb} and in agreement with our numerical results, we will calculate that limit.




\

Calculating the return probability for $\period$-periodic hypercubic spin foams (in the large-$j$ limit) boils down to performing several integrations: on the one hand we have to integrate over spin-foam labels, i.e. $\text{SU}(2)$ spins, on the other hand over the eigenvalues of the Laplacian operator, here captured in branches of frequencies. The number of spins to integrate over increases with growing $\period$, while the number of integrations over eigenvalues remains four for four spacetime dimensions.

We perform the numerical integrations in \url{Julia}\footnote{\url{https://julialang.org/}} using the \url{Cuba}\footnote{See \url{http://www.feynarts.de/cuba/} for the original version and \url{https://github.com/giordano/Cuba.jl} for the \url{Julia} package.} package \cite{Hahn:2005pf}, a set of adaptive algorithms using Monte Carlo techniques suited for higher dimensional integration. Instead of performing all integrations at once, we integrate over the eigenvalues of the Laplacian first. Hence the separate integrations are lower dimensional improving convergence of the algorithms.


\section{Dimensional flow in spin foams}\label{sec:results}


Before we present our results on the spectral dimension of hypercuboid spin foams, in particular how it changes with the scale at which the geometry is probed, we would like to briefly recall the key geometric properties of these configurations.

In the large-$j$ limit hypercuboid spin foams are essentially peaked on flat building blocks that are glued together in a flat way. The spins solely determine the size of faces, the 3-volume of cuboids and the 4-volume of hypercuboids. Thus they affect the entries of the Laplacian, but spacetime in itself remains flat. 

The main effect on the spectral dimension comes from the superposition of geometries of different size, more concretely how geometries of different size are weighted in the path integral depending on the parameter $\alpha$. Indeed this determines which spectral dimension we observe. 
We will first present the results for geometric configurations where each hypercuboid is given by four edge lengths before briefly discussing the spin case.
For all numerical simulations we have chosen the same minimal and maximal cut-off to make the results comparable. For length variables we choose $\lmin = 10^{-3}$ and $\lmax = 10^3$, for spins it is $j_{\text{min}} = 10^{-6}$ and $j_{\text{max}} = 10^6$ accordingly.
We observe qualitatively similar results for various values of periodicity $\period$ justifying the finite-$\period$ computations.  


\subsection{Restriction to geometric configurations}

\subsubsection*{1-periodic configurations}

The simplest example we can study, is the 1-periodic case: the entire geometry is determined by a single hypercuboid and its four edge lengths $l_i$. 
Computing the return probability in this case is rather simple, since the spectrum of the 4D Laplace operator can be computed from the spectra of four 1D Laplace operators (for equilateral 1D lattices) with edge lengths $l_i$. This is straightforward to recognize from \eqref{discrete-laplacian} as the components of the 4D Laplace operator in one direction only depend on the length associated with that direction; the dependence on the other lengths in the 3- and 4-volume cancels, since all angles are right angles. Thus the 1D spectra associated to each of the four dimensions are
\begin{equation}
    \spec_i(l_i,k_i) = \frac{2}{l_i^2} (1 - \cos(k_i)) \quad , \, k_i \in (-\pi,\pi] \quad .
\end{equation}
The full 4D spectrum is the sum of 1D spectra in each direction such that the heat trace factorizes and is solved analytically \cite{Thurigen:2015uc},
\[
P(\tau;\{l_i\}) =  \prod_{j=1}^4 \left ( \int_{-\pi}^{\pi} \d k_j \, \exp \left( -\frac{2 \tau}{l_j^2} (1 - \cos(k_j)) \right) \right)
= \prod_{j=1}^4 \left ( 2\pi\, \e^{-\frac{2 \tau}{l_j^2}} I_0\left(\frac{2\tau}{l_j^2} \right) \right)
\]
where $I_0$ denotes the modified Bessel function of first kind.

While the heat trace takes this simple product form, the amplitude $\av$ does not factorize into a product of amplitudes associated to a single dimension. Hence the integration over length variables must be performed simultaneously, 
\begin{align}
    \langle P(\tau) \rangle_\alpha & = \frac{1}{Z} \int \prod_{i=1}^4 \, \d l_i \; \av(\{l_i\}) \; P(\tau;\{l_i\})  
     = \frac{1}{Z} \int \prod_{i=1}^4 \, \d l_i \; \av(\{l_i\}) \; \prod_{j=1}^4 \left (2\pi\, \e^{-\frac{2 \tau}{l_j^2}} I_0\left(\frac{2\tau}{l_j^2} \right) \right) \quad .
\end{align}
We perform the remaining integral over lengths numerically. The spectral dimension $\Ds(\tau)=\Ds(\tau,\alpha)$ is again defined as the logarithmic derivative 
\begin{equation}
    \Ds(\tau) = -2 \, \frac{\partial \ln \langle P(\tau) \rangle_\alpha}{\partial \ln \tau} \quad .
\end{equation}

\begin{figure}[h]
    \centering
    \includegraphics[scale=0.6]{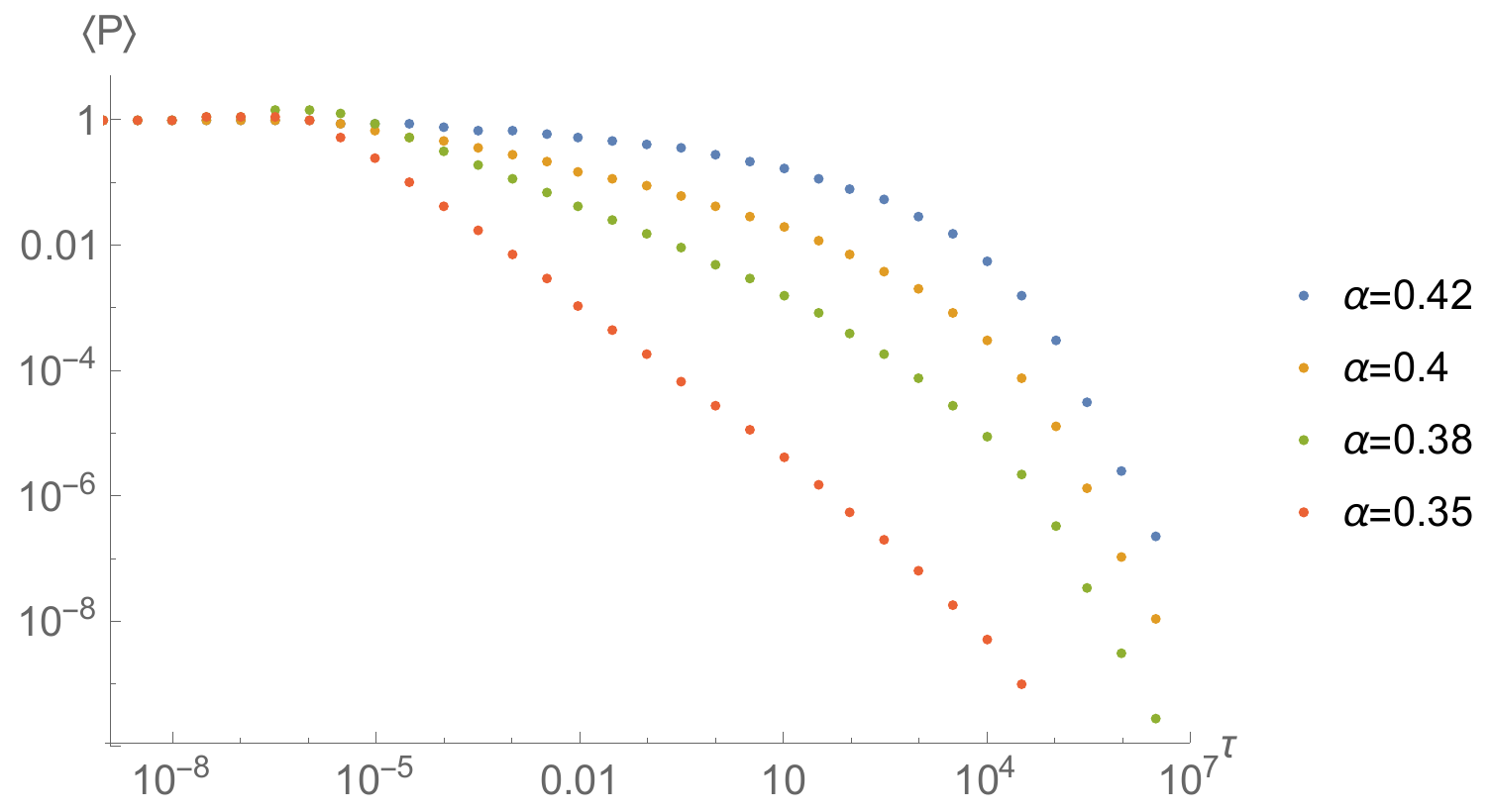}
    \label{fig:RP-1-per-length}
    \includegraphics[scale=0.55]{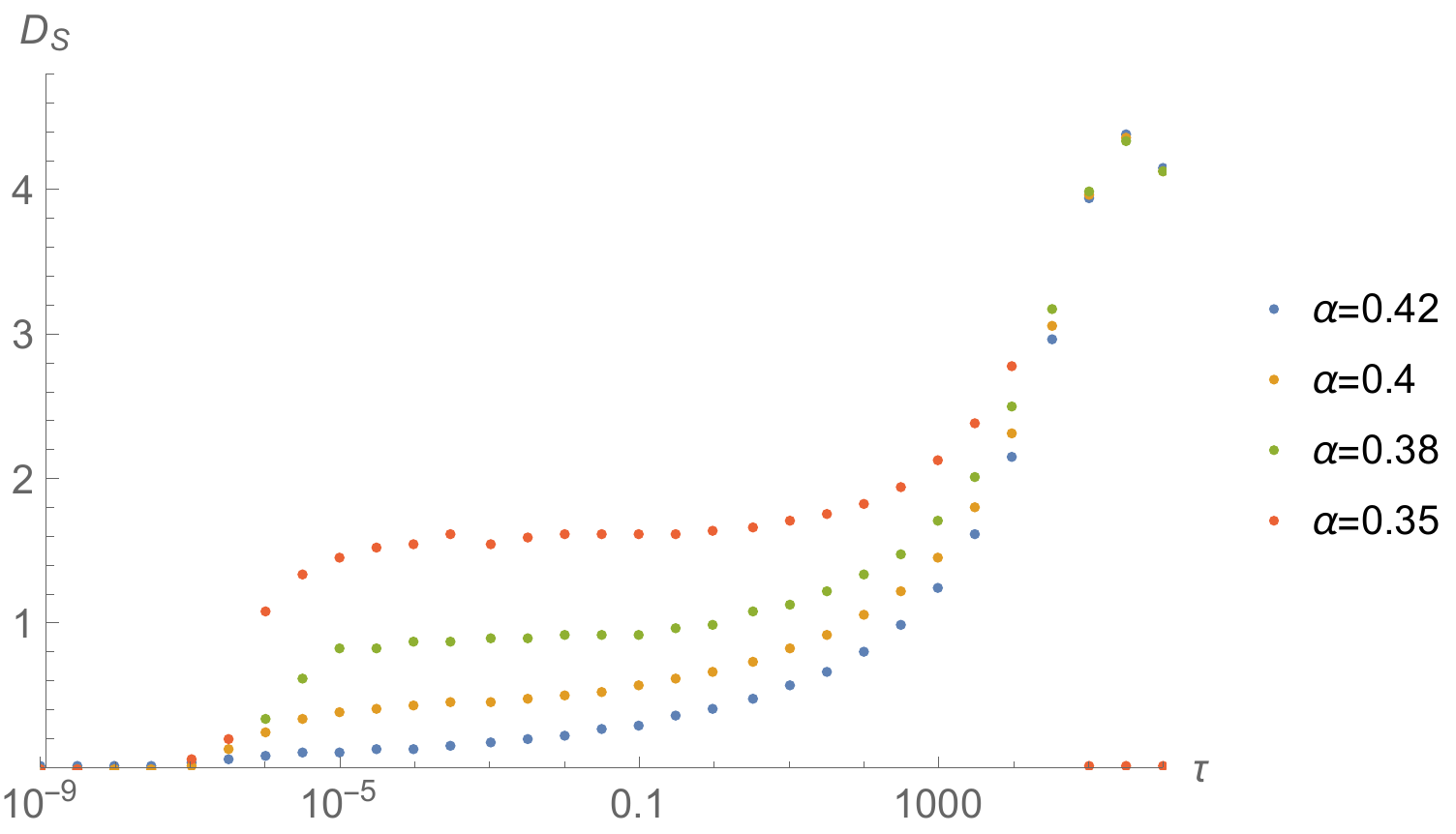}
    \caption{Numerical results of $\langle P(\tau) \rangle$ (left) and $\Ds$ (right) for periodicity $\period=1$ for various values of $\alpha$. The individual points of $\Ds$ are calculated by taking the difference quotient of $\langle P(\tau) \rangle$.}
    \label{fig:1-per-lengths}
\end{figure}

In Fig.~
\ref{fig:1-per-lengths} we plot respectively the return probability and the spectral dimension as a function of $\tau$ over several orders of magnitude for different values of $\alpha$. Table \ref{tab:1periodic} contains the results (with error estimates) from finding the best linear fit for $\ln \langle P(\tau)\rangle_\alpha$ for the middle plateau. Let us describe some general features of the results:


For $\tau \ll \lmin^2$, we only find $\Ds = 0$ as we are probing spacetime below (the smallest allowed) lattice scale. Seen from a random walker, this just means that the random walker was not able to leave the starting 4-cell at all and is thus unable to probe spacetime, hence the return probability is constantly one. For $\tau \gg \lmax^2$, we always find $\Ds=4$. This is straightforward to explain as well because we are probing a superposition of 4D lattices at a scale where {\it all} these geometries possess $\Ds = 4$. As we do not add any geometries of larger lattice scale, we can only find $\Ds =4$ if we increase $\tau$ further. Thus we omit results for $\tau > 10^6$.

The behaviour of $\Ds$ for $\lmin^2 < \tau < \lmax^2$ depends crucially on $\alpha$, namely there is an interval $[\alpha_{\text{min}},\alpha_{\text{max}}]$ in which $\Ds$ changes continuously from $\Ds = 4$  at $\alpha_{\text{min}}$ and $\Ds=0$ at $\alpha_{\text{max}}$. Outside this interval $\Ds = 0$ for $\alpha > \alpha_{\text{max}}$  and  $\Ds = 4$ for $\alpha < \alpha_{\text{min}}$. It is difficult to find $\alpha_{\text{min}}$ and $\alpha_{\text{max}}$ 
numerically; below we will provide an analytical derivation for these parameters, which also estimates them remarkably well in more complicated cases. We comment on the implications of this result later on.

\begin{table}[]
    \begin{tabular}{|  c | c | c | c | c |}
\hline
 $\alpha$    & $0.42$  &  $0.4$      &  $0.38$     & $0.35$ \\ \hline
 $\Ds$        & $0.$    & $0.50 \pm 0.02$ &  $0.97 \pm 0.02$ & $1.64 \pm 0.02$ \\
 \hline
\end{tabular}
\caption{\label{tab:1periodic} Values of spectral dimension $\Ds$ in the 1-periodic case for the plateau $\lmin^2 < \tau < \lmax^2$ obtained from a linear fit to $\langle P(\tau)\rangle_\alpha$.
}
\end{table}

A brief note on the cut-offs is in order: if we remove the minimal cut-off, we do not observe a drop to $\Ds=0$ to small scales unless $\alpha > \alpha_{\text{max}}$. Instead the plateau we observe extends to smallest scales. Similarly if we remove the maximal cut-off we do not observe an increase to $\Ds=4$ for large $\tau$ unless $\alpha < \alpha_{\text{min}}$.


\subsubsection*{2-periodic configurations}

%
%

The next logical step is to study a $2$-periodic spin foam, that is a spin foam completely prescribed by a block of $16$ spin-foam vertices glued together. In the geometric case that we are studying here, each vertex amplitude depends on four edge lengths which are identified across the amplitudes via the glueing. As one would expect, the block of $16$ vertex amplitudes then depends on eight edge lengths in total, two associated to each dimension. 
The scaling properties of each individual vertex amplitude are the same as before, such that the entire amplitude $\Acal$ 
scales according to \eqref{eq:geometric-scaling} as follows (under uniform scaling of all edge lengths):
\begin{equation} \label{eq:2per-scaling}
    \Acal^{(\alpha)}_{\period=2} \propto \prod_v \av(\{\lambda l_i\}) = \lambda^{16(24 \alpha -14)} \prod_v \av(\{ l_i\}) \quad .
\end{equation}

The Laplace operator is slightly more intricate than in the $1$-periodic case, but crucially it still possesses the factorization property across dimensions. To see this let us consider one of its non-zero components in the $x$-direction,
\begin{equation} \label{eq:laplace-2per}
   - \Delta_{x,x+1} = \frac{1}{V^{(4)}} \frac{V^{(3)}}{l^*_x} =
   \frac{1}{l^{(1)}_x l^{(1)}_y l^{(1)}_z l^{(1)}_t} \frac{l^{(1)}_y l^{(1)}_z l^{(1)}_t}{\frac{1}{2} (l^{(1)}_x + l^{(2)}_x)} = \frac{2}{l^{(1)}_x (l^{(1)}_x + l^{(2)}_x)} \quad .
\end{equation}
As in the previous cases, the components of the Laplace operator in one direction only depend on the edge lengths in that particular direction. For the geometric cuboid configurations this generalizes to arbitrary periodicity $\period$. So, the spectrum of the Laplace operator can be computed again via computing the spectra for four 1D lattices of periodicity $\period$, which consist of $\period$ eigenvalues. 

The Laplace operator of a $2$-periodic 1D lattice comes with two off-diagonal entries, here (for example in the $x$-direction)
\[
w_1=\frac{2}{l^{(1)}_x (l^{(1)}_x + l^{(2)}_x)} \quad \text{and} \quad w_2 = \frac{2}{l^{(2)}_x (l^{(2)}_x + l^{(1)}_x)} \, .
\]
This can be seen by swapping $l^{(1)}$ for $l^{(2)}$ in \eqref{eq:laplace-2per}. The spectrum is then derived by Fourier transform exploiting the periodicity of the lattice,
\begin{equation}
\omega^2_\pm(k) = w_1+w_2 \pm \sqrt{w_1^2 +w_2^2 + 2w_1w_2 \cos (k)} \, , \quad k \in (-\pi,\pi] \quad .
\end{equation}
Due to the $2$-periodicity the spectrum has two branches. The branch $\omega_{-}$ goes to zero for $k \rightarrow 0$.

\begin{figure}[h]
    \centering
    \includegraphics[scale=0.6]{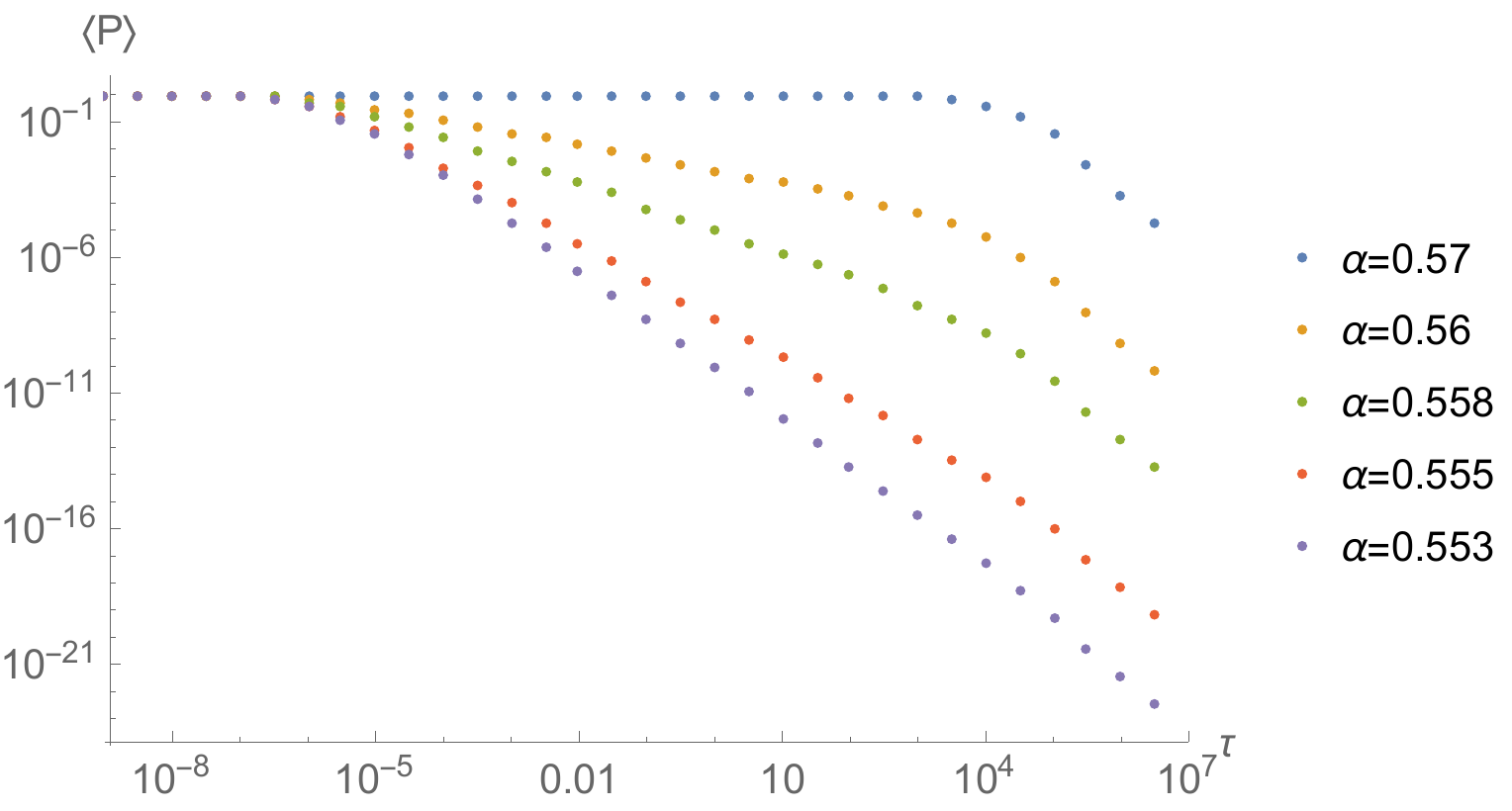}
    \includegraphics[scale=0.55]{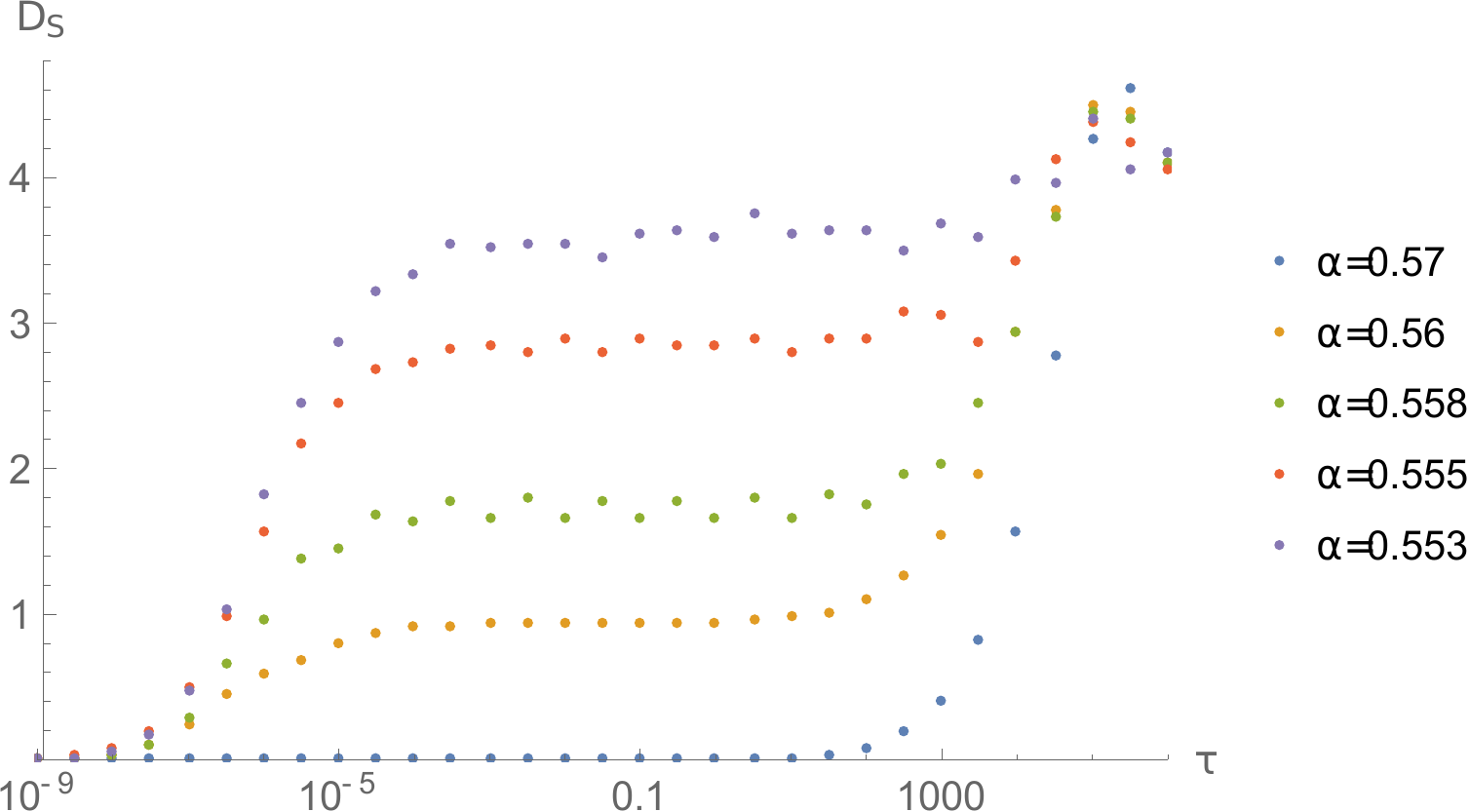}
    \caption{Numerical results of $\langle P(\tau) \rangle$ (left) and $\Ds$ (right) in the 2-periodic case for several values of $\alpha$.}
    \label{fig:2periodic}
\end{figure}

Having defined all necessary ingredients we compute $\langle P(\tau) \rangle_\alpha$ in the $2$-periodic case,
\begin{equation}
    \langle P(\tau) \rangle_\alpha = \frac{1}{Z} \int \prod_{i=1}^8 \d l_i \; \Acal^{(\alpha)}_{\period=2}(\{l_i\}) \; \prod_{j=1}^4 \left( \sum_{s \in \{+,-\}} \int \d k_j \exp\left(-\tau \omega_s(k_j) \right) \right) \quad .
\end{equation}
Analogous to the $1$-periodic case we find the same qualitative features, in particular an interval in $\alpha$ for which we can produce any value of $\Ds$ in $[0,4]$. We present the numerical results in Fig.~
\ref{fig:2periodic} and Table \ref{tab:2periodic}.
\begin{table}[]
    \begin{tabular}{| c | c | c | c | c | c |}
\hline
 $\alpha$   & $0.57$  & $0.56$   &  $0.558$      &  $0.555$     & $0.553$ \\ \hline
 $\Ds$      &  $0.$   & $0.928 \pm 0.011$    & $1.74 \pm 0.02$ &  $2.85 \pm 0.02$ & $3.57 \pm 0.02$ \\
 \hline
\end{tabular}
\caption{\label{tab:2periodic} Values for the spectral dimension $\Ds$ in the 2-periodic case for the plateau $\lmin^2 < \tau < \lmax^2$.
}
\end{table}

While the qualitative features are the same, we observe a change in the $\alpha$-dependence of the spectral dimension. 
First, the interesting interval in $\alpha$ has shifted towards larger values in $\alpha$. 
Second, the size of the interval is considerably smaller compared to the $1$-periodic case. Both effects can be explained by the fact that in the $2$-periodic case the geometry is weighted by $16$ vertex amplitudes, which results in a more ``spiked'' scaling behaviour \eqref{eq:2per-scaling} compared to \eqref{eq:geometric-scaling} in the $1$-periodic case. 
Since the occurrence of a spectral dimension $0<\Ds<4$ crucially depends on the superposition of discrete geometries, the suitable range of $\alpha$ shrinks. Furthermore this changed scaling behaviour also pushes the interval to larger $\alpha$. 

\subsubsection*{3-periodic case}

We conclude the numerical study of geometric quantum cuboids with the 3-periodic case. This case is fairly analogous to the 2-periodic case, but it is already considerably more costly in computational costs.

The spin foam is determined by 81 spin-foam vertices, whose geometry is prescribed by 12 edge lengths. The scaling of this collection of amplitudes changes accordingly.
The spectrum of the Laplace operator is still computed from four 1D spectra, which we compute again via a Fourier transform. Its characteristic polynomial has three solutions $\omega_a^2(k_j)$, i.e. three branches
\footnote{
For $\period\ge3$ the off-diagonal Laplacian entries 
$w_{ij}/V_i$ consist of a symmetric part $w_{ij}=V_{ij}/{l}^\star_{ij}$ with 3-volumes $V_{ij}$ and dual length ${l}^\star_{ij}$ and the inverse 4-volume $V_i$ (note that this Laplacian matrix is still symmetrizable \cite{\COTa}).
For $\period=3$ the spectrum $x=\omega_a^2(k)$ is then given by the three real solutions to the cubic equation
\[
0 = x^3 + \left(\frac{w_{12}+w_{13}}{V_1} +\frac{w_{12}+w_{23}}{V_2} + \frac{w_{13}+w_{23}}{V_3} \right) x^2
+ \frac{V_1+V_2+V_3}{V_1 V_2 V_3} \left(w_{12} w_{13} + w_{12} w_{23} + w_{13} w_{23} \right) x 
+ 2 \frac{w_{12}+w_{13}+w_{23}}{V_1 V_2 V_3} \left( 1 - \cos(k)\right) \, .\nonumber
\]}.
Thus the return probability is computed as follows:
\begin{equation}
    \langle P(\tau) \rangle_\alpha = \frac{1}{Z} \int \prod_{i=1}^{12} \d l_i \; \Acal^{(\alpha)}_{\period=3}(\{l_i\}) \; \prod_{j=1}^4 \left( \sum_{a=0}^2 \int \d k_j \exp\left(-\tau \omega_a(k_j) \right) \right) \quad .
\end{equation}

\begin{figure}[h]
    \centering
    \includegraphics[scale=0.58]{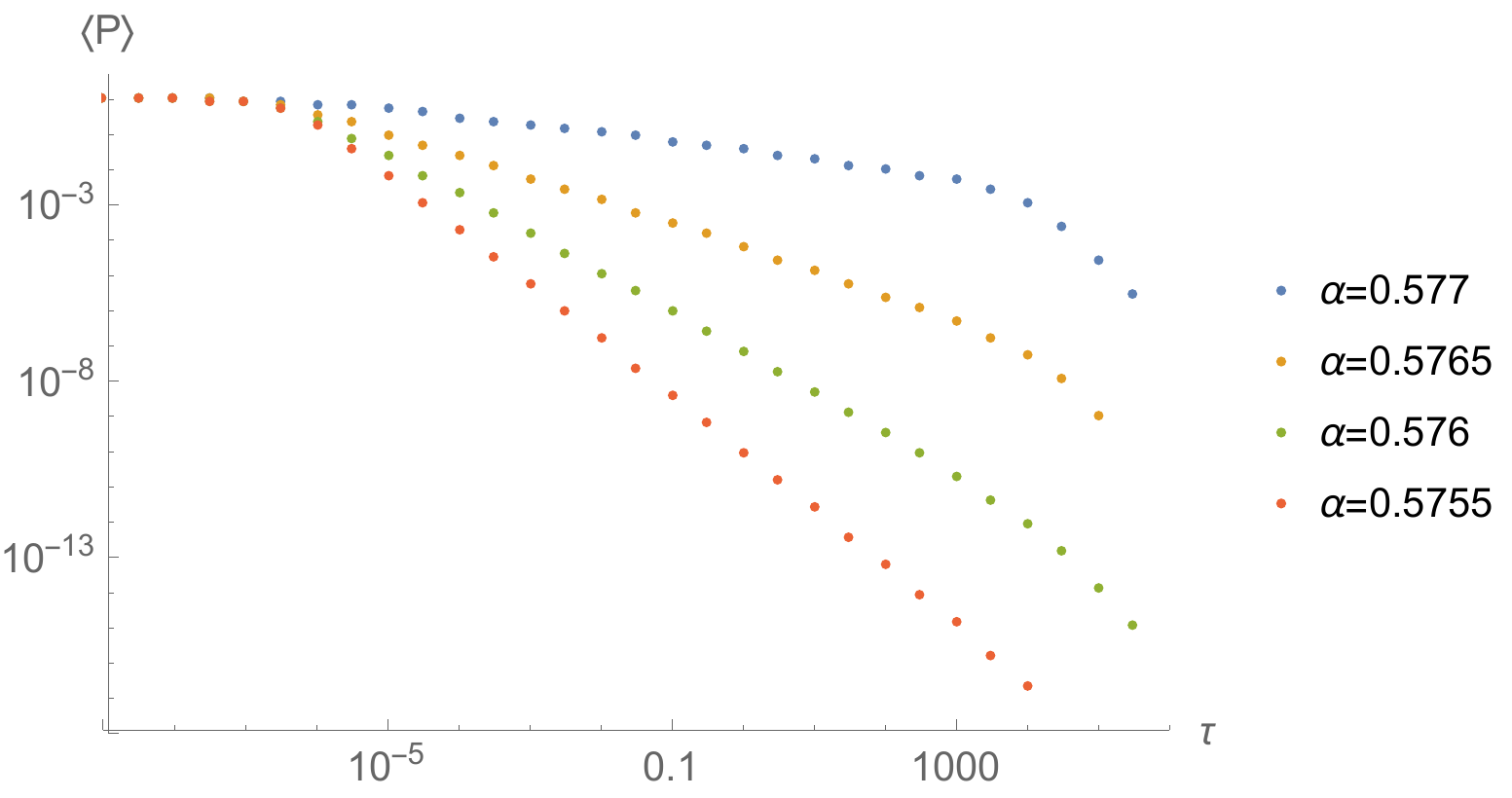}
    \includegraphics[scale=0.55]{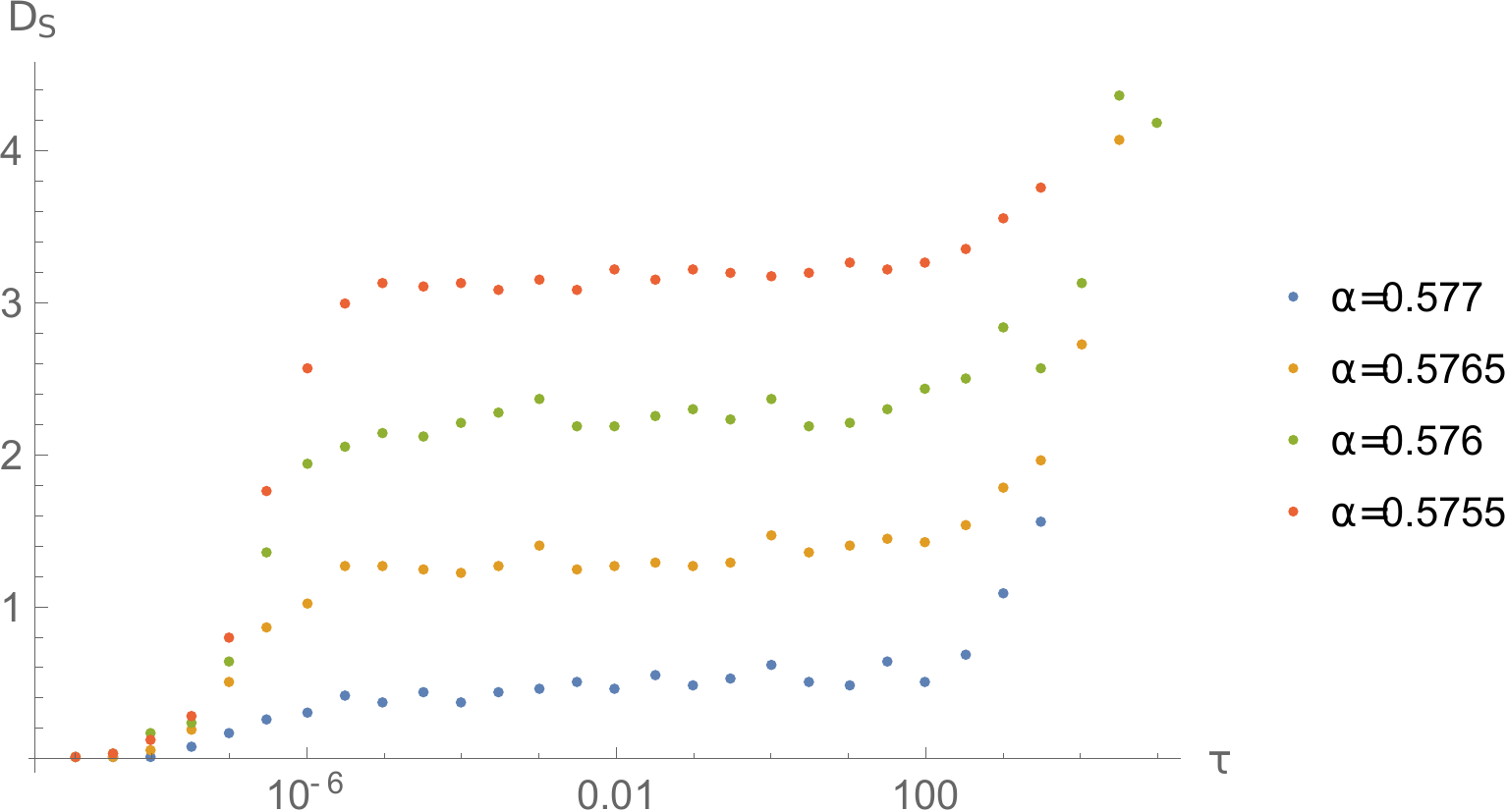}
    \caption{Numerical results of $\langle P(\tau) \rangle$ (left) and $\Ds$ (right) in the 3-periodic case for several values of $\alpha$.}
    \label{fig:3periodic}
\end{figure}

The numerical results are presented in Fig.~
\ref{fig:3periodic} and Table \ref{tab:3periodic}. They confirm the observations from the 1- and 2-periodic cases: Under increasing the periodicity the $\alpha$-interval in which $\Ds$ changes continuously (for the plateau $\lmin^2 < \tau < \lmax^2$) between $0$ and $4$ gets significantly smaller and it is moved to larger values of $\alpha$. However the shift of the interval in $\alpha$ from 2- to 3-periodic is significantly smaller than from 1- to 2-periodic. This indicates that the $\alpha$ interval might shrink and converge to a specific value in the $\period \rightarrow \infty$ limit, which we will discuss later on. 

\begin{table}[]
    \begin{tabular}{| c | c | c | c | c |}
\hline
 $\alpha$ & $0.577$   & $0.5765$  & $0.576$   &  $0.5755$      \\ \hline
 $\Ds$   & $0.50 \pm 0.01$    &  $1.40 \pm 0.03$   & $2.27 \pm 0.02$    & $3.17 \pm 0.02$ \\
 \hline
\end{tabular}
\caption{\label{tab:3periodic} Values for the spectral dimension $\Ds$ in the 3-periodic case for the plateau $\lmin^2 < \tau < \lmax^2$.
}
\end{table}

Since the results only change quantitatively under increasing the periodicity $\period$ we do not expect qualitative changes. For this reason and the growing computational costs, we refrain from studying periodicities $\period > 3$ numerically. Later we will give an analytical argument that this is indeed not necessary.
Before that 
we will briefly study the spectral dimension in the spin case.



\subsection{Spectral dimension of spin configurations}

The restriction to purely geometric configurations is a strong simplification of spin foams. Indeed it is an intriguing question whether the non-geometric configurations can leave an imprint on the qualitative behaviour of the spectral dimension. At this point we would like to remind the reader that it is natural to expect a quantitative difference since the scaling behaviour of spin configurations \eqref{eq:nongeometric-scaling} differs from the one of geometric configurations \eqref{eq:geometric-scaling}.

\begin{figure}[h]
    \centering
    \includegraphics[scale=0.6]{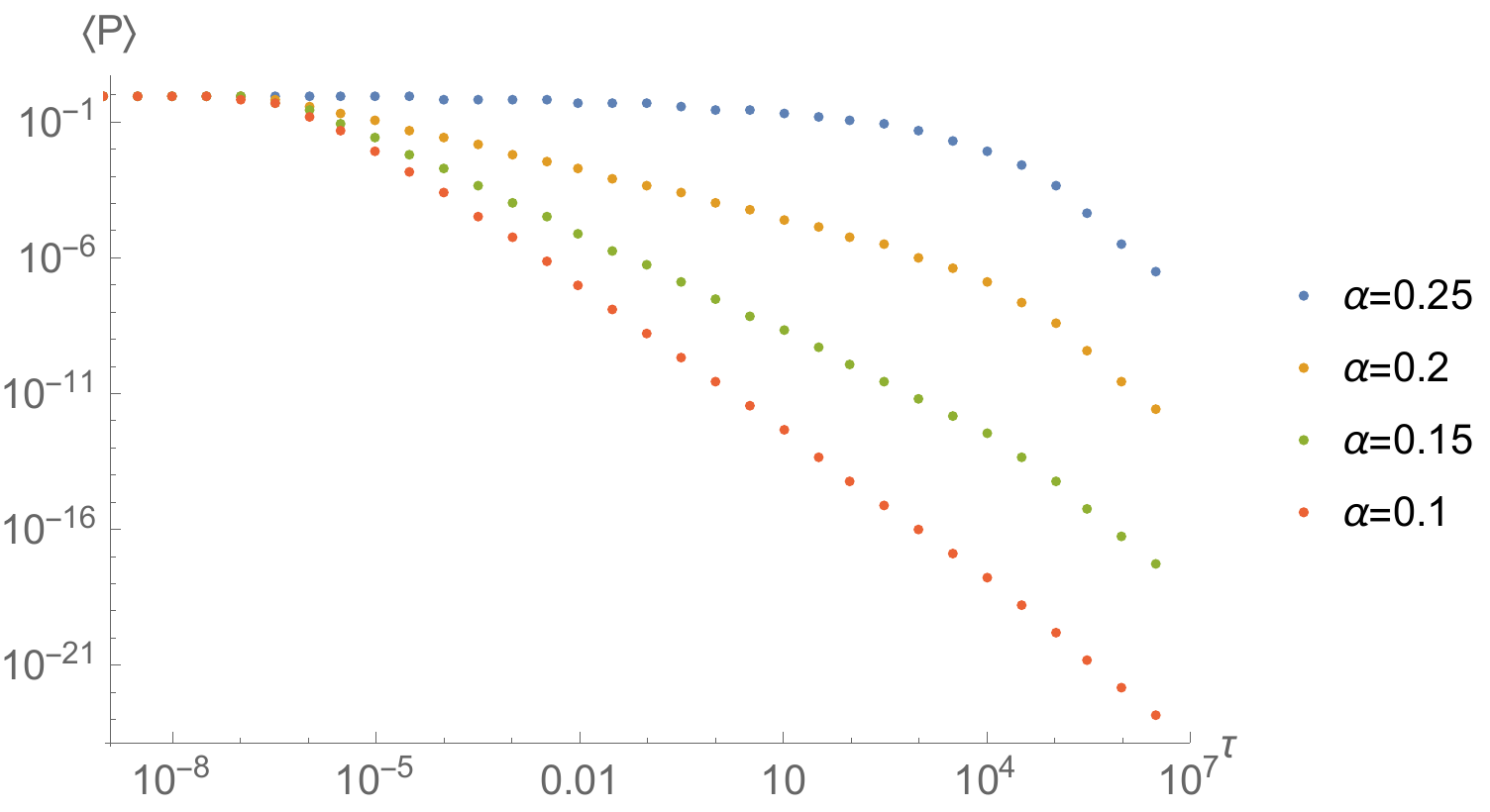}
    \includegraphics[scale=0.55]{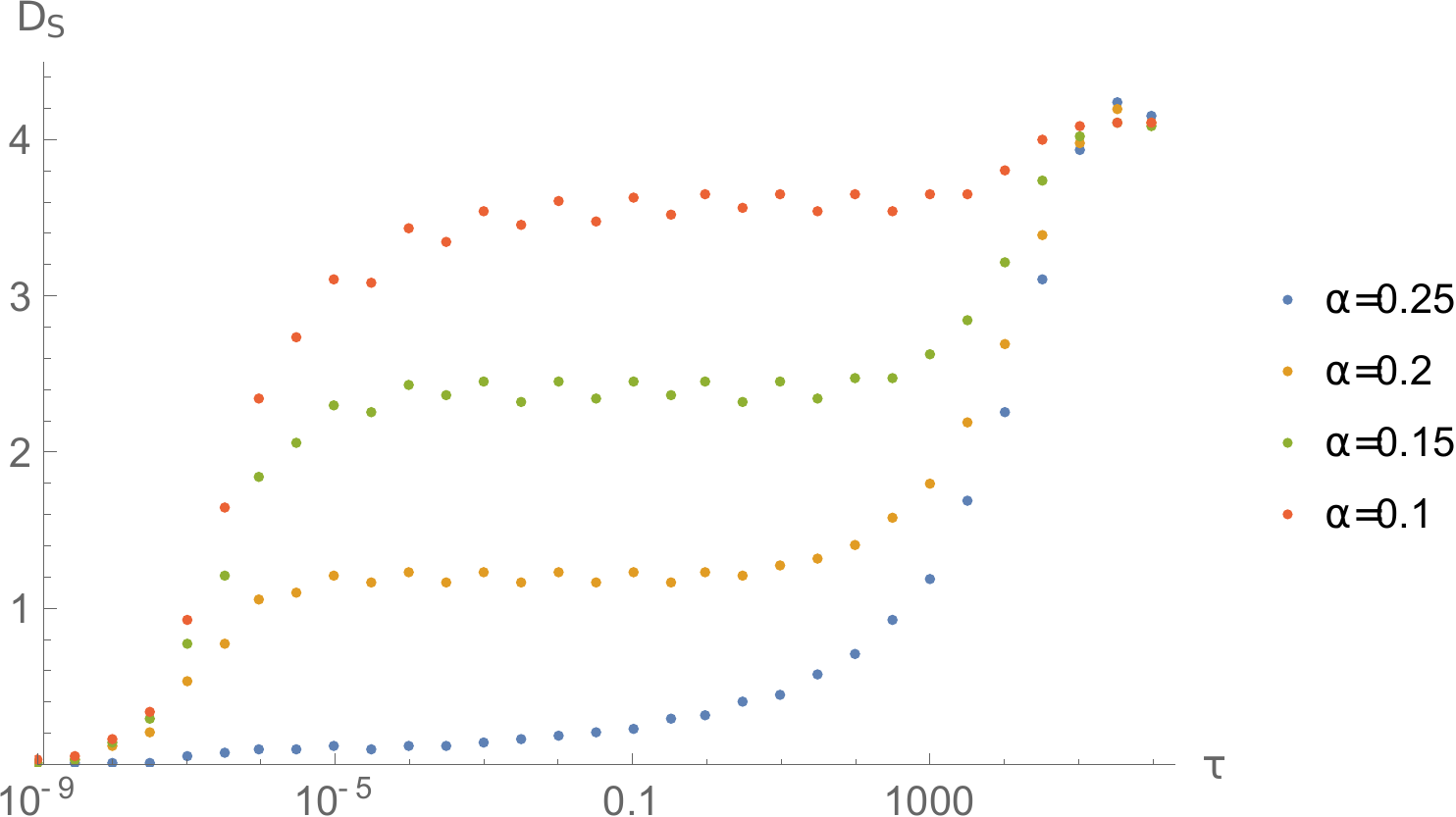}
    \caption{Numerical results of $\langle P(\tau) \rangle$ (left) and $\Ds$ (right) in the 1-periodic case in spin variables for several values of $\alpha$.}
    \label{fig:1periodic_spin}
\end{figure}

As for the geometric case, we start our analysis with the 1-periodic case. This is prescribed by a single hypercuboid given by six spins. Not surprisingly, the spectrum of the Laplace operator for spin configurations cannot generically be expressed any more by the spectra of 1D Laplace operators, since the variables and weights do not factorize any more per dimension. Fortunately, due to the 1-periodicity of the configuration, the components of the Laplacian remain unchanged under translations in any direction.
Hence we compute the spectrum of the 4D Laplace operator yet again from the spectra of four 1D operators with spin dependent weights, e.g. in the $t$-direction,

\begin{equation}
    \spec_t(\{j_m\},k_t) = \frac2{\ell_\gamma^2} \, \frac{(j_1 \, j_2 \, j_3)^\frac{1}{3}}{(j_4 \, j_5 \, j_6)^\frac{2}{3}}
 (1 - \cos(k_t)) \quad , \, k_t \in (-\pi,\pi] \quad ,
\end{equation}
where $j_4$, $j_5$ and $j_6$ also span the $t$-direction, whereas $j_1$, $j_2$ and $j_3$ are purely `spatial'. Thus the calculation of the return probability is analogous to the geometric case, but with a slightly more intricate dependence of the spectrum on the geometry,
\begin{equation}
    \langle P(\tau) \rangle_\alpha = \frac{1}{Z} \int \prod_{i=1}^6 \, \d j_i \; \av(\{j_i\}) \; \prod_{m=1}^4 \left ( \int\, \d k_m \, \exp \left( -\tau \, \spec_m(\{j_i\},k_m) \right) \right) \quad .
\end{equation}
We summarize the results in Fig.~
\ref{fig:1periodic_spin} and Table \ref{tab:1periodic_spin}.

\begin{table}[]
    \begin{tabular}{| c | c | c | c | c | c |}
\hline
 $\alpha$   & $0.3$  & $0.25$   &  $0.2$      &  $0.15$     & $0.1$ \\ \hline
 $\Ds$      &  $0.$   & $0.121 \pm 0.006$    & $1.23 \pm 0.02$ &  $2.42 \pm 0.02$ & $3.53 \pm 0.02$ \\
 \hline
\end{tabular}
\caption{\label{tab:1periodic_spin}  Values for the spectral dimension $\Ds$ in the 1-periodic case with spin variables for the plateau $\lmin^2 < \tau < \lmax^2$.
}
\end{table}

Qualitatively we observe a similar behaviour to the geometric case: For $\tau < j_{\text{min}}$ we only find $\Ds = 0$ while for $\tau > j_{\text{max}}$ we observe only $\Ds = 4$. For diffusion times $\tau$ in between we again observe that there exists an interval $[\alpha_{\text{min}},\alpha_{\text{max}}]$ in which we find plateaus constant in $\Ds$, whose value continuously increases as $\alpha$ is decreased. That is, $\Ds = 0$ for $\alpha > \alpha_{\text{max}}$ whereas $\Ds = 4$ for $\alpha < \alpha_{\text{min}}$. Hence we only observe a quantitative difference to the geometric case, since this $\alpha$-interval is found for significantly smaller $\alpha$. However the inclusion of non-geometric configurations does not seem to leave an imprint on the spectral dimension, at least not in the 1-periodic case.





For periodicity $\period > 1$ the spectrum of the Laplace operator does not split in directions. Thus, it is necessary to consider the entire 4D operator. It is still most convenient to exploit the periodicity of the system and compute its spectrum after a Fourier transform. E.g. for $\period = 2$ this amounts to calculating all eigenvalues as functions of 24 spins and four momenta. Deriving these formulas analytically in full generality is challenging, not to mention for larger periodicity, but computing the spectrum numerically for fixed values of spins and momenta is efficient. Hence we can in principle continue studying the spectral dimension for larger periodicity, however the numerical cost is high. One reason is the fact that we have to perform the momentum integrations all at once instead of one by one. Moreover, already for $\period = 2$ we have to integrate over 24 spins, which is very costly. Thus we leave this question for future research. Nevertheless, due to the qualitative similarity of the non-geometric results to the geometric ones and the robustness of the latter for larger periodicity, we do not expect vastly different results in the spin case for larger periodicity.


%


\subsection{Analytical explanation of numerical results}

The numerical result of a dimensional flow in cuboid spin foams can be explained in terms of the scaling of the amplitudes and an assumption on the scaling of the Laplacian.

The amplitude of cuboid spin foams scales in all cases with an exponent linear in the model's parameter $\alpha$.
In general, the vertex amplitude is a homogenous function of degree $-(a - b \alpha)$.
The relevant instances here are $(a,b)=(9,12)$ for spin variables $x_i = j_i$ according to \eqref{eq:nongeometric-scaling} 
and $(a,b)=(14,24)$ for the restriction to geometric configurations and a transformation to edge-length variables $x_i = l_i$, \eqref{eq:geometric-scaling}.
Thus, expectation values in such a spin foam with $\vx$ vertex amplitudes are sums over spin configurations weighted by
\begin{eqnarray}\label{eq:amplitude-scaling}
\mathcal{A} \propto \prod_{v} \av(\lambda \vec x) \;=\; \lambda^{-(a - b \alpha)\vx} \prod_{v} \av(\vec x) \,.
\end{eqnarray}

In \cite{\COTc}, a generic dimensional flow 
has been found for such superpositions weighted by a power-function measure $x^{-\gamma}$ in the case of a single variable $x$, summing from some $\xmin$ to $\xmax$. 
If the Laplacian is a power function in this variable,
\[\label{eq:lap-scaling}
\Delta(x) = x^{-2\beta} \Delta\,,
\]
where $\Delta$ is the purely combinatorial Laplacian on the hypercubic lattice,
then one finds a dimensional flow from the topological dimension $\std$ at large scales, $\tau \gg \xmax^{2\beta}$, to a spectral dimension
\[\label{eq:dimensional-flow}
\Ds = \left\{ \begin{array}{rccl}
& 0  & ,  &  \frac{\gamma - 1}{\beta} \le 0 \\
& \frac{\gamma - 1}{\beta}  & ,  &  0 < \frac{\gamma - 1}{\beta} < \std \\
& \std & ,  &  \frac{\gamma - 1}{\beta} \ge \std
\end{array} \right . 
\]
at small length scales $\xmin^{2\beta} \gg \tau \gg \xmax^{2\beta}$.
For  $\tau\ll\xmin^{2\beta}$ there is the usual fall-off to zero due to discreteness. Furthermore, the result of an intermediate dimension does not depend on the step size in the sum over $x$ (as long as it is much smaller than $\xmax$).
This result can be directly applied to the test case of a spin-foam sum restricted to equilateral ($\dof = 1$) configurations with $\gamma=(a - b \alpha)\vx$.

The equilateral result can now be generalized to the case of $\period$-periodic spin-foam configurations based on a scaling assumption for the Laplacian. 
For the spectrum of the Laplacian there is some evidence \cite{Sahlmann:2010bb, Thurigen:2018} that the scaling in a single variable \eqref{eq:lap-scaling}
generalizes to $\period$-periodic lattices in terms of averages.
That is, the assumption is that for a sufficiently large number $\dof$ of independent length variables $l_e$ on edges $e$ (which depends in our setting on $\period$, see \eqref{eq:dof})
\[\label{eq:aver-scaling}
\Delta(l_e) \, \approx \, \aver{l^2}^{-1} \Delta \quad ,
\]
where $\aver{l^2}$ is the average over all squared edge lengths $l_i^2$ of a given configuration,

\[
\aver{l^2} = \frac1 \dof \sum_{e} l_e^2 \quad .
\]
Under this assumption we can explain the numerical results analytically.

It should be emphasized that the scaling of the Laplacian's spectrum, though an assumption, is based on a well-motivated conjecture.
In particular, this conjecture stems from an exact result on the perturbative expansion of the spectrum in the momentum $k$ for any $\dof$ \cite{Sahlmann:2010bb}. For the spectral dimension, only the first (quadratic) order is relevant for obtaining the topological dimension above the discreteness scale. Sub-leading orders merely determine the form of a local maximum around the discreteness scale \cite{Thurigen:2018}.
Thus, the assumption is very meaningful in this context and the fact that the numerical results can be explained in this way further supports the conjecture.

\begin{figure}
    \centering
    \includegraphics[scale=0.6]{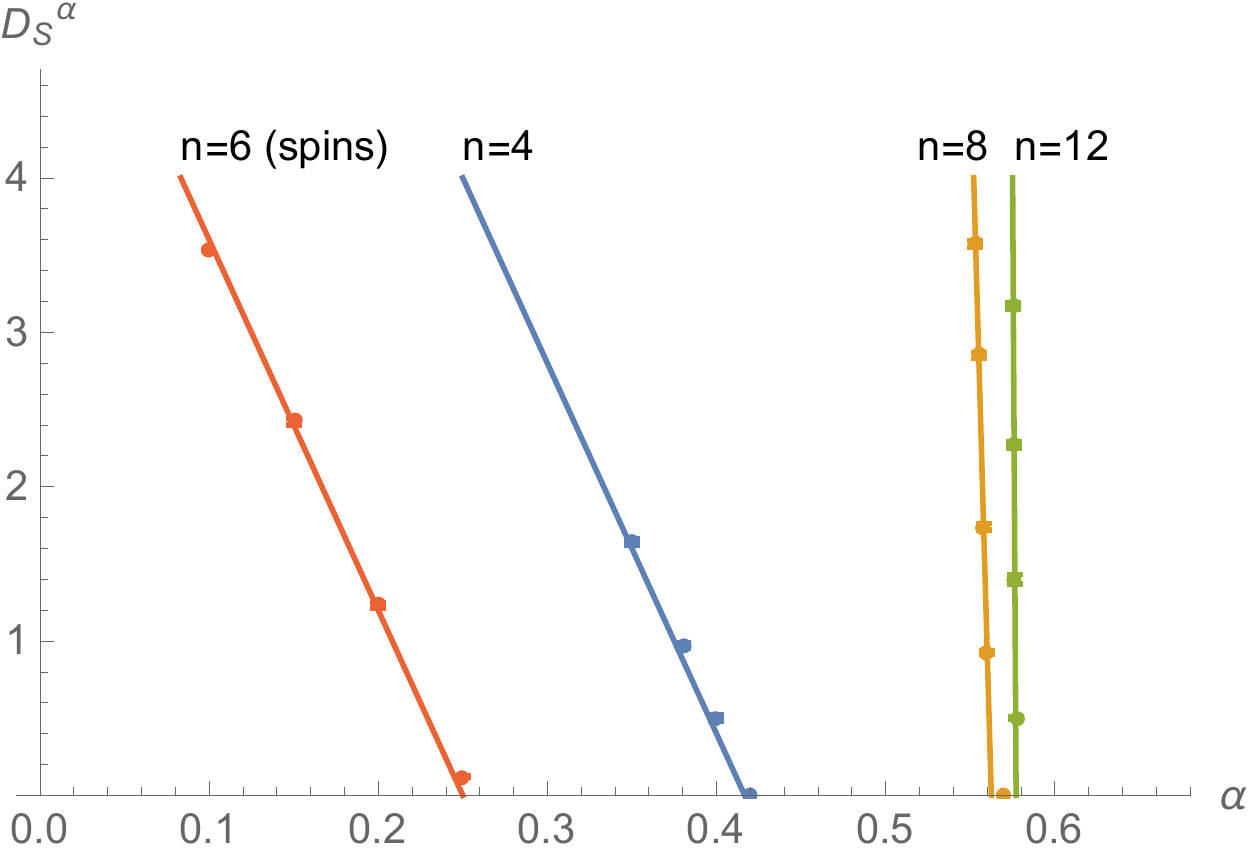}
    \caption{Comparison of numerical results (values and error bars according to Tables I--IV) with the analytical result \eqref{eq:Duv-lengths}, \eqref{eq:Duv-spins} for the intermediate spectral dimension $\Duv$ as a function of the model parameter $\alpha$ for $\dof=4$ ($\vx=1$), $\dof=8$ ($\vx=16$) and $\dof=12$ ($\vx=81$) length variables, as well as $\dof=6$ ($\vx=1$) spin variables. }
    \label{fig:comparison}
\end{figure}

The spin-foam expectation value of the heat trace can now be transformed to a one-dimensional integral, up to an irrelevant overall factor.
We show this first for the restriction to geometric configurations where a transformation to edge-length variables $l_e$ is possible and generalize then to arbitrary spins $j_f$.
The key point is that a transformation of the integral
\[
\expec{P(\tau)} = \frac{\int [\d l_e]^\dof \Acal(l_e) \Tr\, \e^{\tau \lap(l_e)}}{\int [\d l_e]^\dof \Acal(l_e)}
\]
to ``radial" coordinates with radius $\aver{l^2}$ is possible such that only the radial part contributes to the heat kernel expectation value. 
Explicitly, the transformation is
\[
l_e = \sqrt{\dof \aver{l^2}} f_e(\Omega)
\]
where $f_e(\Omega)$ are the standard angular functions in radial coordinates.
In this way, the semiclassical cuboid spin-foam amplitudes factorize into a radial part and angular part $g(\Omega)$ as well,
\[
\Acal(l_e) \propto \prod_{v} \av(l_e) 
= \sqrt{\dof \aver{l^2}}^{-(a - b \alpha)\vx}  g(\Omega) \,.
\]
Then, the angular part in the heat-trace expectation factorizes and 
cancels with the denominator

\begin{eqnarray}
\expec{P(\tau)}_\alpha & = & \frac{\int_{\lmin}^{\lmax} [\d l_e]^\dof \prod_{v} \av(l_e) \Tr\, \e^{\tau \lap(l_e)}}{\int_{\lmin}^{\lmax} [\d l_e]^\dof \prod_{v} \av(l_e)}
\approx 
\frac{ \int_{\lmin^2}^{\lmax^2} \d \aver{l^2} \, \sqrt{\aver{l^2}}^{\dof-2 -(a - b \alpha)\vx} \Tr\,\e^{\tau \aver{l^2}^{-1} \lap}}
{\int_{\lmin^2}^{\lmax^2} \d \aver{l^2} \, \sqrt{\aver{l^2}}^{\dof-2 -(a - b \alpha)\vx}} \\
&=&
\frac1{{\lmax}^{{\dof-(a - b \alpha)\vx}} - {\lmin}^{{\dof-(a - b \alpha)\vx}}}
\int_{\lmin^2}^{\lmax^2} \d \aver{l^2} \, \sqrt{\aver{l^2}}^{\dof-2 -(a - b \alpha)\vx} \Tr\,\e^{\tau \aver{l^2}^{-1} \lap}  \, .
\end{eqnarray}
For the remaining part one can apply again the result \eqref{eq:dimensional-flow} from \cite{\COTc} , now with $\beta = 1/2$ and $\gamma = (2+(a - b \alpha)\vx -\dof)/2$ such that there is a dimensional flow to 
\[\label{eq:Duv-lengths}
\Duv = \vx(a - b \alpha) - \dof
\]
at $\lmin^2 \ll \tau \ll \lmax^2$ for $0 < \vx(a - b \alpha) - \dof < \std$. 
Note that in this way any value $\Duv\in[0,\std]$ can be obtained for some $\alpha$ for a finite number of degrees of freedom $\dof$ and vertex amplitudes $\vx$.
The range of $\alpha$ where such a flow occurs is given by
\[\label{eq:alpha-edgelengths}
\alpha = \frac1 b\left(a-\frac\dof\vx\right) - \frac{1}{b\vx}\Duv \,.
\]  
Numerical calculations in the various cases are in perfect agreement with the analytically derived linear relationship between $\Duv$ and the theory's parameter $\alpha$, as shown in Fig.~\ref{fig:comparison}. 

It is not obvious, however, how the scaling of the Laplacian generalizes from the case of length variables to area variables, \eqref{eq:area-Laplacian}, and thus spins $j_f$. 
At first, a scaling in $\aver{j_f}$ or in $\aver{j_f^2}$ both seems meaningful.
Here we take our numerical results as input where we find a dimensional flow for the sum over $\dof$ spins $j_f$ (see Fig.~\ref{fig:comparison}) to
\[\label{eq:Duv-spins}
\Duv = 2 \left(\vx(a - b \alpha) - \dof \right) \,. 
\]
This can be explained under the assumption for the Laplacian on spin configurations
\[\label{eq:aver-j-scaling}
\Delta(j_f) = \aver{j^2}^{-1/2} \Delta
\]
which means that $\aver{j^2}=\frac1\dof\sum j_f^2$ is the proper average for the scaling assumption.
Performing the same transformation to radial-type coordinates, now in the space of spin configurations,
\[
\Acal(j_f) \propto \prod_{v} \av(j_f) = (\dof \aver{j^2})^{-(a - b \alpha)\vx/2}  h(\Omega) \,,
\]
one finds the heat-trace expectation value 
\[
\expec{P(\tau)} \propto \int_{\jmin}^{\jmax} [\d j_f]^\dof \prod_{v} \av(j_f) \Tr\, \e^{\tau \lap(j_f)}
\approx \frac1 2 \sqrt{\dof}^\dof \int \d\Omega \, \prod_{v} h_v(\Omega)
\int_{\jmin^2}^{\jmax^2} \d \aver{j^2} \, (\aver{j^2})^{\frac{\dof-2 -(a - b \alpha)\vx}{2}} \Tr\,\e^{\tau \aver{j^2}^{-1/2} \lap}  \,.
\]
Thus, the only difference in the resulting equation for the intermediate dimension $\Duv$ between a spin-foam sum over spins $j_f$ and the restricted spin-foam sum over geometric configurations given by $l_e$ is a factor of two stemming from the different scaling of Laplacian in the squared-variable average, that is $\beta=1/4$ for spins and $\beta=1/2$ for edge lengths.

Accordingly, there is an effect of non-geometricity in this spin-foam model on the spectral dimension, though the underlying reason seems to be the number of degrees of freedom and resulting amplitude scaling rather than an explicitly geometric explanation.
The number of degrees of freedom on the $\period$-periodic configurations of the cuboid spin-foam model with its translation invariance is $\dof = \binom{\std}{2}\period^2$ for spin variables on faces, or $\dof = \std \period$ for length variables on edges.  Together with the different amplitude scaling $(a,b)$, \eqref{eq:amplitude-scaling}, one finds a different $\period$-dependent offset in the linear relation between $\Duv$ and $\alpha$, while the slope is the same in both cases for given $\period$ (see Fig.~\ref{fig:comparison}). Restricting to geometric configurations shifts the range of $\alpha$ at which a dimensional flow occurs to smaller $\alpha$ for $\period\ge2$. 
This is true in particular for the large-$\period$ limit which we will discuss in the following.

\section{Full cuboid-spin-foam sum and renormalization group flow}

Though the analytic equations explaining our numerical results turn out to be rather simple, the physical consequences are far reaching.
In the end we are interested in the large-periodicity limit, that is the full cuboid spin-foam sum.
The analytic \eqref{eq:Duv-lengths} and \eqref{eq:Duv-spins} allow us to take this limit. 
It appears that the range of $\alpha$ where an intermediate dimension $0<\Duv<\std$ is observed shrinks to a point $\{\aflow\}$.
We provide two interpretations for this fact.
Considering the large-periodicity limit as a certain thermodynamic limit, one can argue that $\alpha = \aflow$ is the point in the parameter space of such a model where a phase transition from 0-dimensional to $\std$-dimensional spacetime takes place.
Complementary, one can ask the question whether our results are consistent across lattices of different periodicity. In this sense we discuss the possibility to use $\Duv$ as a condition to define a renormalization group flow for $\alpha$.
We discuss these two possibilities -- which should be emphasized not to be exclusive subsequently -- in the two parts of this section.
In any case, an intriguing aspect is that dynamics, more precisely the amplitude $\av$, at $\aflow$ is invariant under global rescaling, connecting the discussion also with the topic of restoration of diffeomorphism invariance.


\subsection{The large-$\period$ limit and scale invariance \label{sec:limits}}

Our numerical computations are approximate in a twofold way. For explicit calculations one has to fix a finite $\period$ and a finite maximal spin $\jmax$.
To obtain exact results it is necessary to take the limits $\period\rightarrow\infty$ and $\jmax\rightarrow\infty$.
The main advantage of the analytic explanation of the results under the power-law assumption for the Laplacian is that they enable us to take these limits. 

\

Both our numerical and analytical studies reveal a universal behaviour of the spectral dimension for cuboid spin foams. For any choice of finite number of spin-foam vertices $V$ or finite periodicity $\period$ we observed essentially three different regimes depending on the parameter $\alpha$: For large $\alpha$ we observe $\Ds = 0$, for small $\alpha$ $\Ds = \std$ and in between those regions exists a small interval in which $\Ds$ can take any value between $0$ and $\std$ changing continuously with $\alpha$. However, the position and size of this interval in $\alpha$ depend sensitively on the size of spin foam. For growing $\period$ (or $\vx$) the size of this interval shrinks and the upper and lower end increase to approach the value in $\alpha$ where the spin-foam amplitude becomes invariant under global scaling, i.e. $\hat{\mathcal{A}}(\{\lambda j_i\}) = \hat{\mathcal{A}}(\{j_i\})$. 

Thus, the relevant quantity to analyse is the parameter $\alpha=\alpha_\period(\Duv)$ at which a certain intermediate dimension $\Duv$ is observed for $\period$-periodic spin-foam configurations.
When approximating large spin-foam configuration sums by   $\period$-periodic configurations, the details of the limit depend on the specific degrees of freedom.
On a hypercubic spin-foam lattice there are in general $\dof \propto \period^\std$  degrees of freedom. However, because of the translation invariance \eqref{Eq:LatticeSymmetry} due to the restriction to cuboids this reduces to
\[\label{eq:dof}
\dof = c_\std \period^p
\]
where $p=2$ and $c_\std = \binom{\std}{2}$ for a general spin configuration while $p=1$ and $c_\std = \std$ for edge-length variables in the restriction to geometric configurations.
In any case it is meaningful to restrict the dynamics according to the periodicity, that is $\vx = \period^\std$ vertex amplitudes. 
As a result, the generalized version of \eqref{eq:alpha-edgelengths} is
\[\label{eq:alpha-period-relation}
\alpha_\period(\Duv) = \frac a b - \frac{c_\std}b  \period^{-\std+p} - \frac{2\beta}{b}\Duv \period^{-\std}
\]
where $\beta=1/4$ in the general spin case and $\beta=1/2$ in the geometric edge-length case, as discussed before.
While any value $\Duv\in[0,\std]$ can be obtained for some $\alpha$ for a finite number of degrees of freedom,
there is a fine-tuning with the periodicity $\period$. 
The larger the $\period$, the smaller the range of $\alpha$ yielding an intermediate dimension $\Duv$. 
In particular, in the limit $\period \rightarrow \infty$, this $\alpha$-interval shrinks to a point $\apt = a/b$.

\

The discontinuous jump from $\Duv=0$ to $\Duv=\std$ at $\apt$ in the limit $\period\rightarrow\infty$ is reminiscent of a phase transition.
First of all, one can indeed consider the large-$\period$ limit as a thermodynamic limit. It is the limit of a large number of degrees of freedom according to \eqref{eq:dof}.
While in quantum gravity volume becomes a quantum observable itself, its expectation value, being an extensive quantity, is expected to scale with the combinatorial size $\size$ of the lattice underlying the spin-foam configurations, in particular with the number of vertices such that their ratio is fixed. 
For technical reasons, \ie to avoid the compactness effect of the spectral dimension, we have considered $\period \ll \size$ in explicit, finite-$\period$ computations in Sec.~\ref{sec:results}.
In the large-$\period$ limit it is however equally meaningful to consider lattices of size $\size = \period$.
In this sense, $\period\rightarrow\infty$ is indeed a thermodynamic limit.

Certainly, the quantity of spacetime dimension is a suitable order parameter in the broadest sense to describe the properties of different phases of a quantum geometry. 
A different (effective) dimension of a spacetime, in particular, when this spacetime is made up of intrinsically $\std$-dimensional building blocks, implies that it is ordered in a different way.
The spectral dimension $\Ds$ captures a particularly physical aspect of spacetime dimension, \ie the effective dimension as tested by a scalar field. 
Hence the spectral dimension is a highly non-local quantity encapsulating information of the entire spacetime. 
However this is not surprising in theories of quantum gravity which are expected to have non-local features, e.g.~due to diffeomorphism invariance. Thus the definition of physically meaningful local quantities is rather challenging. 
In causal dynamical triangulations, for comparison, it is also common to use averaged geometric quantities such as the number of triangulation vertices per number of 4-cells as order parameters \cite{Ambjorn:2012vc}.
These quantities have a more precise meaning as order parameters since they are conjugate to coupling parameters on the level of the state sum. 
Still, the jump of the spectral dimension from $\Duv=0$ to $\Duv=\std$ at $\alpha=\apt$ is at least a strong sign for a phase transition.

The phase transition to $\std$-dimensional spacetime becomes even more relevant in the large-$\jmax$ limit.
The scale of the dimensional flow from $\std$ to $\Duv\le\std$ is given by the largest length scale $\lmax^2\sim\jmax$ up to which the spin-foam is summed.
However, this is just an artificial cutoff, to be sent to infinity in most spin-foam models%
\footnote{Spin-foam models for quantum gravity with a non-vanishing cosmological constant $\Lambda$ often come with a natural cut-off $\jmax$ on the spins \cite{Smolin:1995vq,Major:1995yz,Borissov:1995cn,Noui:2002ag,Fairbairn:2010cp,Han:2010pz,Turaev:1992hq,Haggard:2014xoa}. One possibility are quantum groups at root of unity, where $\jmax$ is related to the level $k$ of the quantum group.}. 
As a consequence, this flow from $\Duv$ to $\std$ is shifted to infinity, that is, the spectral dimension has the value $\Duv$ on all scales (above the discreteness scale $\jmin$).
In particular, for an $\alpha$ yielding $\Duv<\std$ semiclassical quantum spacetime is indeed of dimension smaller than the observed classical $\std$-dimensional spacetime.
The usual $\std$ dimensions occur only below some minimal $\alpha_{\text{min}}$, either continuously 
for finite $\period$, 
or discontinuously at $\alpha=\apt$ in the thermodynamic limit $\period\rightarrow\infty$. 
In this sense $\std$-dimensional spacetime emerges in this spin-foam model.

\

The critical parameter $\apt$ has a special physical relevance, as it is the point where the amplitude $\av$ becomes invariant under global rescaling, i.e.~only the shape of the cuboid determines the value of the amplitude, not its scale. Indeed we know from \cite{Bahr:2016co, Bahr:2017kr} that $\alpha$ determines whether small or large spins are preferred in the spin-foam state sum, where $\apt$ marks the turning point between these two domains.

This tells us that in the $\period \rightarrow \infty$ limit, the spectral dimension $\Ds$ is solely determined by the global scaling behaviour of the amplitude, where the scale-invariant amplitude marks the transition between the phases characterized by $\Ds = 0$ and $\Ds = 4$. The individual shape of the cuboids does not seem to influence the spectral dimension. Interestingly, this observation strongly resonates with the action of an Abelian subgroup of diffeomorphisms, which transforms a cuboid configuration into another cuboid configuration. Such a diffeomorphism acts by shifting an entire hyperplane in the direction orthogonal to it to obtain another hypercubioid configuration. Hence such a diffeomorphism only changes how flat space is subdivided into flat cuboids, which are glued together in a flat way. 

The independence of the spectral dimension on the shape of the cuboids in the $\period \rightarrow \infty$ limit implies that it is indeed invariant under such diffeomorphisms
which might indicate that diffeomorphism invariance is restored in this limit at $\alpha = \apt$. 
This is further underlined by two previous results on cuboid spin foams restricted to geometric configurations. 
In \cite{Bahr:2016co} two glued hypercuboids were considered under translations of the middle cube along which they are glued while keeping the total volume fixed. It was found that $\av$ is almost invariant under such hyperplane translations for $\alpha \approx 0.6$. 
Similarly in \cite{Bahr:2016dl,Bahr:2017kr} indications for a UV-attractive fixed point of the renormalization group flow of $\av$ were found around $\alpha\approx 0.628$. It has been conjectured that broken diffeomorphism symmetry of discrete theory gets restored at such a fixed point \cite{Bahr:2009mc,Bahr:2011uj}. 
Remarkably both these values are in close proximity to $\apt = {14}/{24} \approx 0.583$ in the present context. The root of the discrepancy might be due to the fact that the calculations in \cite{Bahr:2016co,Bahr:2017kr} were performed for finite lattices, whereas our result holds for $\vx,\period \rightarrow \infty$.

Conversely, we can revert the logic and wonder when to expect a non-trivial spectral dimension. In our study of the spectral dimension for finite periodicity $\period$ of the spin foam, we observed two different regimes in $\alpha$, one with $\Ds = 0$ and one with $\Ds = 4$ separated by a small region in which it changes continuously. In the limit of $\period \rightarrow \infty$ the two regimes persist and we can readily assign a spectral dimension to them. However the intermediate region, which contains any $0 < \Ds < 4$, shrinks to a single point $\apt$. Hence we cannot infer a value of the spectral dimension right on this transition, yet we expect it to be non-trivial there. Moreover, if this is the point on which diffeomorphism symmetry is restored, we conjecture that the spectral dimension shows a non-trivial behaviour there. 

Naturally, this reasoning should be taken with a grain of salt. The spectral dimension is, by definition, a global and non-local observable, in which many properties of a geometry get washed out. This is even more true in the case of a quantum geometry where one 
sums over all geometries allowed by the theories. Nevertheless, the fact that in the $\period \rightarrow \infty$ limit only the scaling properties of the amplitude determine the spectral dimension and microscopic properties do not appear to play any role hints towards a restoration and connection to diffeomorphism invariance.

As an exception, spin-foam configurations on a hypercubic lattice without any symmetry on the variables are a special case with a slightly different limit $\apt$. 
While the cuboid spin-foam model implies translation invariance, our analytical argument also applies to less restricted spin-foam sums. 
If all variables in the spin-foam configurations are independent, that is $p=\std$ in \eqref{eq:dof}, the $\alpha_\period(\Duv)$ relation \eqref{eq:alpha-period-relation} is modified and has the limit
\[\label{eq:alpha-period-full}
\alpha_\period(\Duv) = \frac{a-c_\std} b - \frac{2\beta}{b}\Duv \period^{-\std} \underset{\period\rightarrow\infty}{\longrightarrow} \frac{a-c_\std} b  
\]
corresponding to a shift $a \mapsto a - c_\std$. 
No such spin-foam model with $\dof\propto\period^\std$ variables has been calculated explicitly yet.
But it is obvious that such a model, being much less restricted than the cuboids, would describe also curvature and other degrees of freedom.
One could take this simple argument for a different $\apt$ thus as a hint that such a model would indeed also be scale invariant at such nontrivial $\apt$.


\subsection{Renormalization and the spectral dimension as order parameter}

Complementary to the thermodynamic limit $\period\rightarrow\infty$ one can pose the question of how the results depend on the chosen periodicity $\period$ and whether the results for the spectral dimension are consistent. 
Indeed, spin foams of different periodicity correspond to different choices of discretizations. Generically, for the same $\alpha$, we then obtain different values for the spectral dimension for different periodicities $\period$.
To make this consistent, we can revert the reasoning and ask which $\alpha$, i.e.~which spin-foam amplitudes, must we assign to spin foams of different periodicity in order to observe the same spectral dimension. In this sense, we use the value of the spectral dimension as a criterion to define a renormalization group flow for different periodicities, namely for finer and coarser discretizations.
If we then consider the flow in the refining direction, i.e.~growing $\period$, we reobtain the large-$\period$ limit and observe the flow $\alpha \rightarrow \aflow$ for all $0 < \Ds < 4$. 
In this sense we can interpret $\aflow$ as the UV fixed point of the renormalization group flow.

\

In a nutshell, the idea behind renormalization in spin foams is to model the same physical transition on different discretizations. This is done by relating and identifying boundary states across Hilbert spaces by embedding maps and looking for consistent dynamics \cite{Dittrich:2014ui,Dittrich:2012ba,Dittrich:2013xwa,Dittrich:2012he,Dittrich:2016dc}. This naturally extends to expectation values of observables, which should be the same for different discretizations. Conversely one can invert this logic and define a renormalization group flow by requiring that expectation values of observables agree \cite{Bahr:2016co, Bahr:2016dl, Bahr:2017kr}. The spectral dimension, which is a global observable defined for all discretizations, is a possible candidate (with certain limitations\footnote{For the spectral dimension to show meaningful behaviour, one cannot choose the underyling discretization to be too small. Otherwise one only observes the compactness of geometry, see also the discussion in Sec.~\ref{sec:periodic-configurations}.}).

In this article we have studied infinite 4D lattices 
without a boundary, hence the refining formalism via embedding maps does not readily apply. 
As a further restriction we have set the spin foam to be $\period$-periodic to have better control over the number of variables. A natural idea to relate the spectral dimension across such discretizations is to compare an $\period$-periodic spin foam to a $2\period$-periodic one.

This comparison works conceptually as follows: The $2\period$-periodic spin foam is regarded as the refinement of the $\period$-periodic one. This implies that each of the repeated cells is subdivided once in each dimension, resulting in $2^\std$ times more lattice sites for the $2\period$-periodic spin foam. 
For this comparison to be reasonable, we must ensure that we compare the spectral dimension for similar (superpositions of) geometries.

For the cuboid spin foams studied here, the only relevant geometric quantities are the spins on faces, or the lengths on edges in the geometric restriction.
Thus our results so far are labelled by the minimal and maximal allowed spins $\jmin, \jmax$ or edge lengths $\lmin, \lmax$.
In order to compare $\period$- and $2\period$-periodic lattices, their total minimal and maximal scales must agree. 
For geometric cuboids this naturally implies that the minimal and maximal lengths in the $2\period$-periodic case are just half the size of their respective counterparts in the $\period$-periodic case. 
Accordingly, for spins which relate to areas the refinement step yields
\[
\jmin \mapsto \frac1{\sqrt2}\jmin \quad, \quad \jmax \mapsto \frac1{\sqrt2}\jmax \,.
\]
In this sense the $2\period$-periodic configuration is a refinement of the $\period$-periodic one via a rescaling.


The comparison of the spectral dimension across spin foams of different periodicity $\period$ leads to a flow in $\alpha$ as follows:
%
%
From the analytical explanation of our numerical results we have an explicit formula for the intermediate spectral dimension $\Duv = \Duv(\alpha,\period)$ as a function of $\alpha$ and $\period$, via the number of vertices $\vx=\period^\std$ and number of degrees of freedom $\dof = c_\std \period^p$ with $0<p\le \std$ covering all the possible cases discussed in Sec.~\ref{sec:limits}.
If $\Ds$ is supposed to be the same under refining $\period_{i+1} = 2 \period_i$, we have to assign specific parameters $\alpha_i=\alpha_{\period_i}$ and $\alpha_{i+1}=\alpha_{\period_{i+1}}=\alpha_{2\period_i}$ to each periodicity. Then we interpret the $\period_i$-periodic spin foam for $\alpha_i$ as the effective, coarse-grained amplitude of the $\period_{i+1}$-periodic spin foam for $\alpha_{i+1}$, both giving rise to the same spectral dimension $\Duv$.  According to \eqref{eq:Duv-lengths} and  \eqref{eq:Duv-spins} the parameters $\alpha_i$ have to satisfy
\begin{eqnarray}\label{eq:Duv-refinement}
2\beta \Duv(\alpha_i,\period_i)
=\left(a - b \alpha_i\right)\period_i^{\std} - c_\std \period_i^p
= 2\beta \Duv(\alpha_{i+1},\period_{i+1})
& = & \left(a - b \alpha_{i+1}\right)\period_{i+1}^{\std} - c_\std \period_{i+1}^p \\
& = & \left(a - b \alpha_{i+1}\right)2^\std \period_i^{\std} - c_\std 2^p \period_i^p \nonumber
\end{eqnarray}
or equivalently
\[\label{eq:flow-equation}
(a - b \alpha_{i+1}) - 2^{-\std}(a - b \alpha_i)
= c_\std 2^{-\std} (2^p -1) \period_i^{p-\std} \, .
\]
For $p<\std$ the right-hand side vanishes for large $\period$.
The equation is then solved by
\[
\alpha_i = \frac{a}{b} - \frac1{2^{\std(i-1)}} \left(\frac{a}{b} - \alpha_0\right)
\underset{i\rightarrow\infty}{\longrightarrow} \frac{a}{b}
\]
which converges to $\aflow=a/b$ after many refinement steps, independent of some initial parameter value $\alpha=\alpha_0$.
Taking the $\period$-dependent right-hand side of \eqref{eq:flow-equation} into account, the more general flow depending on $\period_i = 2^i\period_0$ is
\[
\alpha_i = \frac{a}{b} - \frac1{2^{\std(i-1)}} \left(\frac{a}{b} - \alpha_0\right)
- \frac1{\period_i^{\std-p}} \left(1-\frac1{2^{\std(i-1)}}\right)\frac{2^p -1}{2^\std -1}\frac{c_\std}{b}
\]
In particular, one observes that for a full set of independent variables, $p=\std$, and only in this case, the fixed point of the flow is shifted to $\aflow=(a-c_\std)/b$, in agreement with \eqref{eq:alpha-period-full}.

Remarkably, this result does not depend on the scaling of the Laplacian captured by $\beta$.
Furthermore, the details of the coarse graining are not important either. Any refinement $\period_{i+1} = \kappa \period_i$, $\kappa>1$, leads to the same fix point $\aflow$ (where formally $\kappa$ might even be real).
The kinds of variables of the models captured by $p$ and $c_\std$, in particular length or spin variables, make a difference only if $p=\std$, that is, if the number of variables are proportional to the number of vertex amplitudes.

There is subtlety regarding the consistency of the flow equations \eqref{eq:flow-equation}.
For a given value of intermediate dimension $0<\Duv<\std$, there exists only one value $\alpha$ for any (finite) periodicity producing this spectral dimension. 
Thus, the renormalization group flow $\alpha_i \rightarrow \alpha_{i+1}$ is unambiguous and actually invertible. 
However, for $\Duv = 0$ or $\Duv = 4$ we usually find large domains in $\alpha$ giving such a spectral dimension. Consequently when considering the flow, no unambiguous flow can be defined directly. This quite significantly restricts the applicability of \eqref{eq:flow-equation} as the $\alpha$ interval permitting $0<\Duv<\std$ shrinks rapidly.

Nevertheless, one can extend the flow to all $\alpha$ covering also the regime of $\Ds = 0$ and $\Ds = 4$.
On these values it is never going to flow out of this phase by the renormalization group flow. Hence we can define a new $\alpha_i$ under renormalization for it by considering the minimal value $\alpha_{i}(\Duv=0)$ for $\Duv = 0$ as well as the maximal value $\alpha_{i}(\Duv=4)$. 
If we have $\Duv = 0$, the $\alpha_i$ to be renormalized is  $\alpha_i\geq \alpha_i(0)$, whereas for $\Ds = 4$, the corresponding $\alpha_i \leq \alpha_i(4)$. In this sense, we can extend \eqref{eq:flow-equation} to the entire domain of $\alpha$.

At this stage, a comment on the flow itself is in order. Given the previously mentioned extension, we always see a flow under refinement of $\alpha \rightarrow \aflow$. In that sense the flow is UV attractive. However, this statement should be taken with a grain of salt. It clearly holds that if we start, for a given $\period$, with an $\alpha$ leading to a $0 < \Ds < 4$, yet outside this interval, which quickly shrinks for growing $\period$, we have defined the flow to be the same as inside the interval.
This behaviour might indicate that the spectral dimension is not an ideal observable to define a renormalization group flow, but still serves as a good order parameter for identifying different geometric phases of the model. 

Note that this renormalization group flow is defined under the assumption that the Laplace operator - and thus the way the scalar field probes spacetime - does not change under this flow, which is well motivated by regarding the scalar field as a mere test field. However, in general {\it both} matter and gravity need to be renormalized at the same time, e.g.~to describe how matter effectively propagates on an effective spacetime. Furthermore it is necessary to identify consequences of choosing a particular discrete Laplace operator. We hope to shed more light on these intriguing questions in future research.


\

\section{Conclusions} \label{Sec:Conclusion}

In this work we have calculated the spectral dimension of flat quantum spacetime as defined by the restriction of the EPRL-FK spin-foam model to cuboid geometries.
More precisely we studied $\period$-periodic spin foams, both numerically and analytically, and found the following general behaviour:
The spectral dimension vanishes below the minimal scale $\jmin$ and flows to $\Ds=4$ above a maximal scale $\jmax$. In between we have found an intermediate value $0\le\Duv\le4$ that depends sensitively on the parameter $\alpha$ characterizing the face amplitude of the spin-foam model as well as the number of degrees of freedom $\dof$ 
parametrized by the periodicity $\period$. For larger $\alpha$, we always find $\Duv = 0$, whereas $\Duv = 4$ for small $\alpha$. 
In between, for finite $\period$, there exists an interval in $\alpha$ in which $\Duv$ increases linearly with decreasing $\alpha$. This interval shrinks with increasing $\period$.
Under the assumption that the Laplacian scales with a certain power of the inverse mean square of the spin-foam variables we have analytically derived the relation between $\Duv$ and $\alpha$ for arbitrary periodicity $\period$ which is in good agreement with our numerical results.
This allows to generalize our numerical results to any $\period$.
It furthermore predicts the results for models on a lattice with different variables, e.g.~fewer symmetries.

The analytical results permit us to take the $\period \rightarrow \infty$ limit in which the $\alpha$-interval of intermediate dimensions $\Duv$ shrinks to a point marking a discontinuous transition between $\Ds = 0$ and $\Ds = 4$. We have interpreted this as evidence for a phase transition from 0-dimensional to 4-dimensional spacetime. 
The point of this transition is precisely given by $\alpha = \apt$ on which the spin-foam amplitudes become invariant under global rescaling. Hence the spectral dimension in the $\period \rightarrow \infty$ limit solely depends on the scaling behaviour of the amplitude and not on the shape of the individual cuboids.
This hints towards restoration of (an Abelian subgroup of) diffeomorphisms in this limit.
To the best of our knowledge, this is the first time that such an emergence of 4-dimensional spacetime has been found in the context of spin-foam models.

\

Naturally these results must be taken with a grain of salt: the cuboid spin-foam model we considered here is a restricted version of the EPRL-FK model. We define it on a hypercubic lattice and fix the intertwiners to be of cuboid shape. Thus, in the large-$j$ limit these lattices are essentially flat geometries, which are subdivided in different ways. As a consequence, several features which we expect to influence the spectral dimension are not accessible in this model. 
Intertwiner degrees of freedom, encoding different shapes of 3D building blocks, are not summed over, and thus we cannot encode curvature.
As another consequence, the model is not sensitive to oscillating spin-foam amplitudes. 
Furthermore, 
we have not studied the deep quantum regime by restricting ourselves to the large-$j$ limit, such that log oscillations as proposed to be a general feature of quantum geometry \cite{Calcagni:2017im} are not observed. 
Anyway, on sufficiently large length scales the spectral dimension is not affected by quantum (small-spin) effects since short scale geometries get exponentially suppressed in the return probability for growing diffusion time.
In spite of these restrictions, we expect our results to carry over qualtiatively to more general spin-foam models (for Riemannian signature on semi-classical length scales). 

However, the spectral dimension is very sensitive to the combinatorial structure of a discrete, or discretized spacetime theory.
Beyond regular lattices as exploited here, it remains a huge challenge to find at all ensembles of combinatorially random geometries which are effectively 4-dimensional 
\cite{Bonzom:2016to, Lionni:2017tk}. So far, only severe constraints on the combinatorics of the cell complexes such as a foliation into spatial hypersurfaces as in causal dynamical triangulations \cite{Ambjorn:2005fh, Ambjorn:2012vc}, or combinatorial translation invariance as in the hypercubic lattice are known to lead to a regime of $\Ds=4$.
The use of such lattice is very meaningful in the present context where spin-foam configurations are considered as a discretization of continuum spacetime and the dynamics are eventually defined through coarse graining and a renormalization group flow.
On the other hand, our results have no straightforward generalization to a context where different combinatorial ensembles dominate, for example triangulations dual to melonic diagrams as in tensor models \cite{Gurau:2016wk,Gurau:2012hl, Bonzom:2011cs}. 
Since such triangulations effectively obey the structure of branched polymers \cite{Gurau:2013th}, one would expect also for a spin-foam model with such combinatorics a maximal value of the spectral dimension of $\Ds=4/3$.
In this sense, the expectation that our results still hold on more general spin-foam models applies to a generalization of the variables on a lattice, not to a generalization of the lattice to any other combinatorial dynamics.

Despite these limitations,
our model allows us to isolate one particular aspect of spin foams affecting the spectral dimension, namely the superposition of geometries. 
Indeed, most 
of the single discrete geometries summed over in the path integral have a spectral dimension $D_S = 4$ above their respective effective lattice scale given by the spins.
Hence, one might not be surprised to find a phase with $D_S = 4$ for the quantum geometry. 
However, we have observed that depending on the spin-foam amplitude the quantum geometry is described by $D_S < 4$ or that it might even be 0-dimensional. 
The latter occurs when the amplitude prefers large spins 
while an intermediate value is the effect of a subtle balance of spins of all sizes. 
If a regime with $D_S = 4$ is supposed to appear in general in spin-foam models,
it occurs where this balance tends towards a preference of small spins in the partition function.

Furthermore, in the more general spin case our model allows for non-geometric configurations, which can be interpreted as torsion. However, despite quantitative differences we have not observed a qualitatively different behaviour from the geometric case.
In this way we would also like to see our work as a proof of principle upon which to build future research on. 
One crucial idea for our work is to use periodic configurations. From the numerical perspective it is the indispensable ingredient to build feasible simulations by keeping the number of variables reasonably small. Moreover it allowed us to study the full spectrum of the Laplacian and avoid the issue of compactness of the configurations. We are curious to see whether these ideas are applicable to more scenarios in spin-foam models.

We expect these results to transfer qualitatively also to more general spin-foam models, 
at least for Riemannian signature.
The cuboid restriction of the EPRL model is a restriction to flat spacetime on the level of the quantum state sum.
An obvious next step in this line of research would be to compute the spectral dimension on less restricted models which cover curvature degrees of freedom, for example the frustrum model \cite{Bahr:2017bn} with cosmological constant \cite{Bahr:2018vv}.
Still, our guess would be that local curvature excitations do not effect the qualitative behaviour of the spectral dimension as a global observable. 
Already in the cuboid model computed here there are torsion-like degrees of freedom due to the non-geometricity. We have seen that their main effect as compared to the restriction to geometric configurations is only a quantitative modification to the $\alpha$-dependence of the result, which is eventually due to the different number of degrees of freedom.
We expect similar modifications when adding more local degrees of freedom such as curvature.

Extending our study to models for Lorentzian signature is more challenging for several reasons. 
One is simply that the large-spin asymptotics for the Lorentzian model is known only partially and the case of cuboid restriction remains to be worked out.
Another reason is that, 
if one considers the field-propagation as actual physical, the possibility to return to the same point in spacetime in the Lorentzian context would imply closed-timelike curves or random walkers propagating back in time. 
This issue could be addressed transferring the definition of causal spectral dimension as studied in causal sets \cite{Eichhorn:2017djq} to spin foams. 
On the contrary, one could argue that a notion of dimension should not directly depend on the causal structure.
In fact, the definition of the quantum spectral dimension depends only on the spectral properties of the geometry (implicit in the Laplacian) and the quantum dynamics as captured by the spin-foam amplitude. 
The causal structure is then already induced by the spectrum of Lorentzian Laplacians and Lorentzian amplitudes.

\acknowledgements
S.St. thanks Lisa Glaser for encouraging him to study the spectral dimension in spin foams. 
J.T. thanks Benjamin Bahr and the II. Institute for Theoretical Physics at the University of Hamburg for hospitality at the initial stage of this work.
The authors thank Benjamin Bahr, Lisa Glaser, Bianca Dittrich, Sylvain Carrozza, Lee Smolin, Renate Loll, Jan Ambj\o rn, Timothy Budd, Astrid Eichhorn, Sumati Surya, Frank Saueressig and Hanno Sahlmann for enlightening discussion during various stages of this work.

S.St. was supported in part by the Perimeter Institute for Theoretical Physics. Research at Perimeter Institute is supported by the Government of Canada through Innovation, Science and Economic Development Canada and by the Province of Ontario through the Ministry of Research, Innovation and Science. 
S.St. was further supported by the project BA 4966/1-1 of the  German  Research  Foundation  (DFG).

J.T. was supported by the German Academic Exchange Service (DAAD) with funds from the German Federal Ministry of Education and Research (BMBF) and the People Programme (Marie Curie Actions) of the European Union's Seventh Framework Programme (FP7/2007-2013) under REA grant agreement n$^\circ$ 605728 (P.R.I.M.E. - Postdoctoral Researchers International Mobility Experience).



\bibliographystyle{JHEP}
\bibliography{main}

\end{document}